\documentclass[english,11pt,a4paper]{article}
\usepackage[T1]{fontenc}
\usepackage[latin9]{inputenc}
\usepackage{multirow}
\usepackage{amsmath}
\usepackage{amsthm}
\usepackage{amssymb}
\usepackage{graphicx}

\makeatletter

\providecommand{\tabularnewline}{\\}

\numberwithin{equation}{section}

\usepackage{jheppub}
\usepackage{slashed}
\usepackage[yyyymmdd]{datetime}
\usepackage{xcolor}
\definecolor{fade}{RGB}{200,200,200}

\makeatother

\usepackage{babel}
\begin{document}
\title{ Dark matter produced from right-handed neutrinos }

\abstract{ Right-handed neutrinos (RHNs) provide a natural portal
to a dark sector accommodating dark matter (DM). In this work, we
consider that the  dark sector is connected to the standard model
only via RHNs and ask how DM can  be produced from RHNs. Our framework
concentrates on a rather simple and generic interaction that couples
RHNs to a pair of dark particles. Depending on whether RHNs are light
or heavy in comparison to the dark sector and also on whether one
or both of them are in the freeze-in/out regime, there are many distinct
scenarios resulting in rather different results. We conduct a comprehensive
and systematic study of all possible scenarios in this paper. For
illustration, we apply our generic results to the type-I seesaw model
with the dark sector extension, addressing whether and when DM in
this model can be in the freeze-in or freeze-out regime.  Some observational
consequences in this framework are also discussed. 

 }

\author[a]{Shao-Ping Li} 
\author[a]{and Xun-Jie Xu} 
\affiliation[a]{Institute of High Energy Physics, Chinese Academy of Sciences, Beijing 100049, China} 
\preprint{\today}  \emailAdd{spli@ihep.ac.cn} \emailAdd{xuxj@ihep.ac.cn}

\maketitle

\section{Introduction}

Right-handed neutrinos (RHNs) are a highly motivated extension to
the Standard Model (SM), often considered to be responsible for the
generation of neutrino masses~\cite{Minkowski:1977sc,yanagida1979proceedings,GellMann:1980vs, glashow1979future,mohapatra1980neutrino}
and the matter-antimatter asymmetry of the universe~\cite{Fukugita:1986hr,Davidson:2008bu}.
It is thus tempting to ask whether they are related to dark matter
(DM), the existence of which has been evidenced by a variety of cosmological
and astrophysical observations~\cite{Bertone:2004pz}.  

Perhaps the simplest answer is that RHNs themselves could be DM.
Indeed, given an appropriate mass ($\sim$keV) and small mixing, 
RHNs have long been considered as a popular DM candidate, known as
sterile neutrino DM (see Refs.~\cite{Dasgupta:2021ies,Kusenko:2009up,Abazajian:2017tcc,Adhikari:2016bei}
for reviews).  Typically being produced via the Dodelson--Widrow~\cite{Dodelson:1993je}
or Shi--Fuller~\cite{Shi:1998km} mechanism, sterile neutrino DM
is not absolutely stable, albeit long-lived compared to the age of
the universe. Its decay could cause observable X-rays which, together
with recent Lyman-$\alpha$ observations \cite{Baur:2017stq,Irsic:2017ixq,Palanque-Delabrouille:2019iyz,Garzilli:2019qki},
have ruled out the original scenario proposed by Dodelson and Widrow.

Going beyond the simplest answer, RHNs due to their singlet nature
under the SM gauge symmetries could readily couple to a dark sector.
 There has been rising interest in a rather simple interaction, $\nu_{R}\chi\phi$,
which couples a RHN $\nu_{R}$ to a dark fermion $\chi$ and a dark
scalar $\phi$~\cite{Pospelov:2007mp,Falkowski:2009yz,Gonzalez-Macias:2016vxy,Escudero:2016ksa,Tang:2016sib,Batell:2017rol,Bandyopadhyay:2018qcv,Becker:2018rve,Chianese:2018dsz,Folgado:2018qlv,Chianese:2019epo,Hall:2019rld,Blennow:2019fhy,Bandyopadhyay:2020qpn,Hall:2021zsk,Chianese:2021toe,Biswas:2021kio,Coy:2021sse,Coy:2022xfj,Barman:2022scg,Coito:2022kif}.
This is the minimal setup for RHN-portal DM with absolute stability. 

Given the interaction $\nu_{R}\chi\phi$, DM could be produced from
RHNs in the early universe, and if being sufficiently produced, it
could also annihilate and return energy and entropy to the SM thermal
bath via RHNs at a late epoch. The thermal evolution of the dark sector
depends on the abundance of RHNs.  In Refs.~\cite{Coy:2021sse,Coy:2022xfj},
RHNs are produced via freeze-in and dominantly decay to $\chi$ and
$\phi$. The freeze-in is achieved with a tiny Yukawa coupling in
the type-I seesaw Lagrangian. In this setup, the DM relic abundance
depends on the amount of RHN particles being produced,  almost independent
of the dark sector coupling. In Ref.~\cite{Barman:2022scg}, the
type-I seesaw model is also assumed but RHNs are in thermal equilibrium,
which is a natural consequence if the seesaw Yukawa couplings are
not fine tuned. From RHNs to the dark sector, it is a freeze-in process,
assuming the dark sector coupling is sufficiently small. Therefore,
depending on parameter settings, different scenarios (e.g.~freeze-in
from the SM to RHNs versus freeze-in from RHNs to the dark sector)
can be achieved in the same model. 

Since various scenarios might be possible for RHN-portal DM, we aim
to conduct a comprehensive and systematic analysis of all possible
scenarios. We investigate, case by case, the thermal evolution and
relic abundance of DM, assuming that thermal equilibrium may or may
not be reached between the SM and RHNs, and between RHNs and the dark
sector. The criteria for whether the equilibrium can be reached are
derived for all cases. Depending on whether the dark sector particles
are lighter or heavier than RHNs, the production of DM  is dominated
by decay or scattering processes, respectively. The former if kinematically
allowed is usually much more efficient than the latter, leading to
very different results for heavy and light RHNs.  Thus we believe
that a comprehensive investigation to cover various possibilities
is necessary and might be  useful for more extensive studies. 

Despite that many scenarios in this framework have been studied before,
we would like to point out that there are still some scenarios that
have not been considered in the literature and might have interesting
observational consequences. For example, when $\nu_{R}$ is in the
freeze-in regime and lighter than DM, a considerably large amount
of energy budget could be stored via $\nu_{R}$ into the dark sector
and then returned to $\nu_{R}$ at a relatively late epoch. This would
be particularly interesting if neutrinos are Dirac particles because
the returned energy budget could lead to a potentially large contribution
to the effective number of relativistic species, $N_{{\rm eff}}$.

Our work is structured as follows. In Sec.~\ref{sec:Framework} we
introduce the framework of RHN-portal DM and propose four generic
cases to be investigated. In addition, by briefly reviewing the Boltzmann
equation, we also set up the necessary formalism for calculations.
Then in Sec.~\ref{sec:DM-relic}, we present detailed analyses. Readers
solely interested in the results are referred to Tab.~\ref{tab:result1}
for a summary.    For illustration, we apply our results to type-I
seesaw DM in Sec.~\ref{sec:seesaw}. A few  observational consequences
are discussed in Sec.~\ref{sec:Observation}. Finally we conclude
in Sec.~\ref{sec:Conclusion} and relegate some calculations to the
appendix.

\section{Framework\label{sec:Framework}}

\subsection{Lagrangian and conventions}

Our framework assumes that the dark sector is connected to the SM
content only via a RHN, $\nu_{R}$.  As aforementioned, the minimal
setup for RHN-portal DM with absolute stability requires a pair of
dark sector particles, a fermion $\chi$ and a scalar $\phi$. The
Lagrangian reads:
\begin{align}
\mathcal{L}\supset y\chi\nu_{R}\phi+\text{h.c.},\label{eq:Yukawa}
\end{align}
where $y$ is a  Yukawa coupling.  Throughout we adopt the Weyl
spinor notation for all fermions so the product of $\chi$ and $\nu_{R}$
is  simply written as $\chi\nu_{R}$. The particle masses of $\nu_{R}$,
$\chi$, and $\phi$ are denoted by $m_{\nu_{R}}$, $m_{\chi}$, and
$m_{\phi}$, respectively. For simplicity, we assume 
\begin{equation}
m_{\chi}<m_{\phi}\thinspace,\label{eq:-48}
\end{equation}
so that only $\chi$ is a DM candidate while $\phi$ produced in the
early universe would eventually decay to $\chi$.  Since in many
models the dark sector may have a symmetry responsible for the DM
stability, we assume $\phi$ is a complex scalar. Switching to a real
scalar would change some of our results by a factor of $1/2$.

For all particles considered in this work, we assume that there is
no asymmetry in the thermal dynamics between particles and anti-particles:
\begin{equation}
n_{X}=n_{\overline{X}}\ \ (\text{for\ }X=\chi,\ \phi,\ \text{\ensuremath{\nu_{R}}},\ \nu_{L},\ \cdots)\thinspace,\label{eq:-23}
\end{equation}
where $\overline{X}$ denotes the antiparticle partner of $X$ and
$n_{X(\overline{X})}$ denotes the number density of $X$ ($\overline{X}$),
respectively. Under the assumption  of Eq.~\eqref{eq:-23}, in this
work we use notations of particles and anti-particles  interchangeably.
For instance, the process $\nu_{R}+\overline{\nu_{R}}\to\chi+\overline{\chi}$
will be written as $2\nu_{R}\to2\chi$ for brevity. In case of potential
confusions, one can always recover the notations of particles and
anti-particles explicitly.

\subsection{Four generic cases of the RHN portal}

\begin{figure}
\centering

\includegraphics[width=0.5\textwidth]{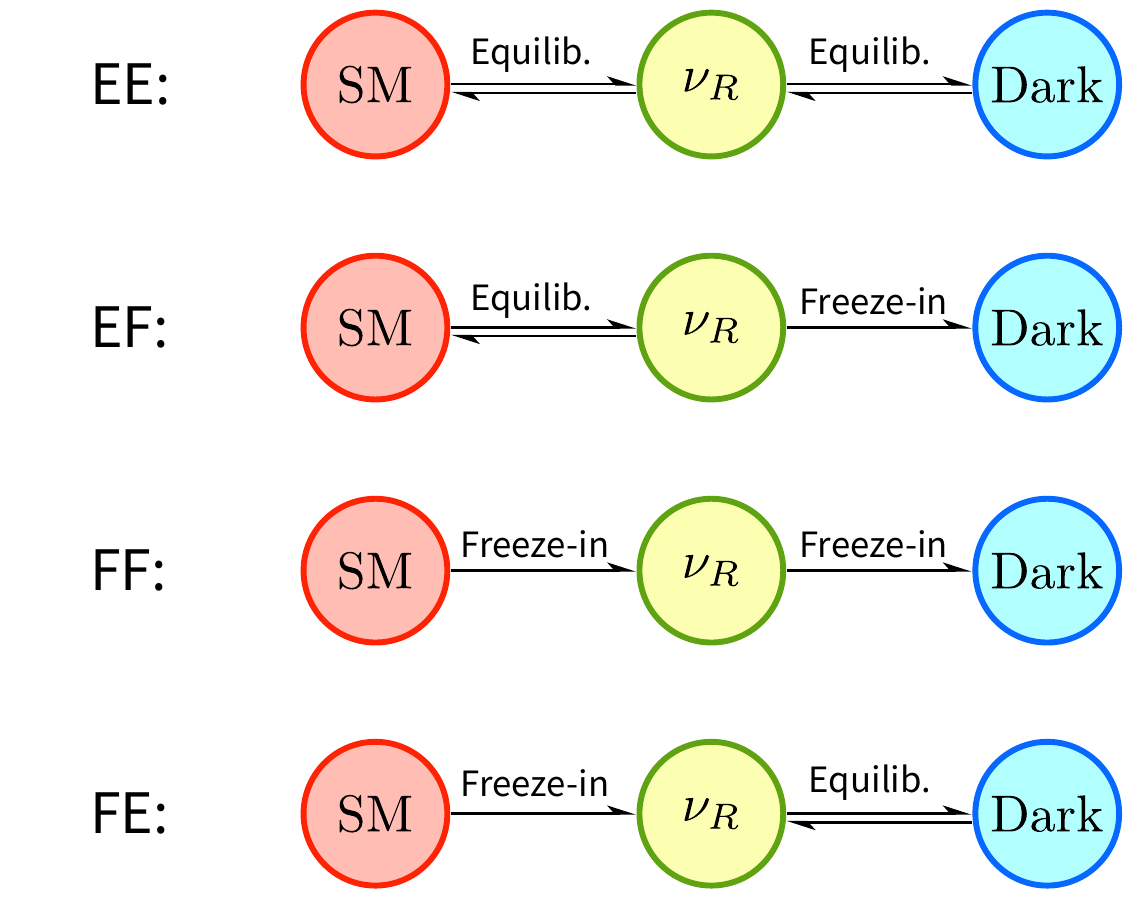}

\caption{Four different cases for the entropy transfer among the SM thermal
bath, $\nu_{R}$, and the dark sector. \label{fig:4-cases}}
\end{figure}

We aim at a comprehensive study of various possible scenarios arising
from  the framework of Eq.~\eqref{eq:Yukawa}. Depending on the coupling
$y$ and also on how RHNs interact with the SM, there are four possible
cases we may encounter, as illustrated in Fig.~\ref{fig:4-cases}
and elucidated as follows.
\begin{itemize}
\item Equilibrium-Equilibrium (EE): both $\nu_{R}$ and the dark sector
are in thermal equilibrium with the SM thermal bath. In this case,
if $\nu_{R}$ is lighter than DM, then it is essentially in the standard
WIMP paradigm. The DM abundance is determined by the moment when $2\chi\to2\nu_{R}$
freezes out. If $\nu_{R}$ is heavier than DM, it becomes more complicated
since $2\chi\to2\nu_{R}$ is kinematically suppressed in the non-relativistic
regime.  
\item Equilibrium-Freeze-in (EF): $\nu_{R}$ is in thermal equilibrium with
the SM whereas the dark sector is frozen-in from $\nu_{R}$. In this
case, the DM abundance is determined by the integrated production
rates of $\chi$ and $\phi$ from $\nu_{R}$. The production rates
crucially depend on whether $\nu_{R}\to\chi\phi$ is kinematically
allowed or not. In any case, the production process $2\nu_{R}\to2\chi$
is always kinematically allowed at high temperatures. 
\item Freeze-in-Freeze-in (FF): no thermal equilibrium is reached between
$\nu_{R}$ and the SM thermal bath, or between the dark sector and
$\nu_{R}$.  DM is produced via a two-step freeze-in mechanism. 
\item Freeze-in-Equilibrium (FE): $\nu_{R}$ is frozen-in from the SM and
keeps thermal equilibrium with the dark sector. In this case, the
DM abundance depends on how many $\nu_{R}$ particles have been produced
through the thermal history, if $\nu_{R}$ is heavier than $m_{\chi}+m_{\phi}$
and dominantly decays to $\chi$ and $\phi$.  If $\nu_{R}$ is lighter
than $m_{\chi}$, the DM abundance also relies on the freeze-out of
$2\chi\to2\nu_{R}$.   
\end{itemize}

\paragraph{}

\subsection{Boltzmann equations}

For a generic species $X$, the number density $n_{X}$ is governed
by the following Boltzmann equation:
\begin{equation}
\frac{dn_{X}}{dt}+3Hn_{X}=C_{X}\thinspace,\label{eq:-1}
\end{equation}
where $H$ is the Hubble parameter and $C_{X}$ is the collision term
including contributions of all processes that create or annihilate
$X$ particles. For $X=\nu_{R}$, $\chi$,  or $\phi$ in our framework,
each collision term may receive contributions from several reaction
processes:  
\begin{align}
C_{\chi} & =C_{2\nu_{R}\leftrightarrow2\chi}+C_{\nu_{R}\leftrightarrow\chi\phi}+C_{\phi\leftrightarrow\chi\nu_{R}}+C_{2\phi\leftrightarrow2\chi}\thinspace,\label{eq:-3}\\
C_{\phi} & =C_{2\nu_{R}\leftrightarrow2\phi}+C_{\nu_{R}\leftrightarrow\chi\phi}+C_{\chi\thinspace\nu_{R}\leftrightarrow\phi}+C_{2\chi\leftrightarrow2\phi}\thinspace,\label{eq:-4}\\
C_{\nu_{R}} & =C_{{\rm SM}\leftrightarrow\nu_{R}}+C_{2\chi\leftrightarrow2\nu_{R}}+C_{2\phi\leftrightarrow2\nu_{R}}+C_{\phi\leftrightarrow\chi\nu_{R}}+C_{\chi\phi\leftrightarrow\nu_{R}}\thinspace.\label{eq:-5}
\end{align}
Here the  subscripts of the $C$'s on the right-hand side indicate
the specific processes, except for ${\rm SM}\leftrightarrow\nu_{R}$
which generically stands for unspecified processes connecting the
SM and $\nu_{R}$. In practice, not all the processes above have to
be taken into account; some may be suppressed or kinematically forbidden.

For a generic two-to-two process, $1+2\leftrightarrow3+4$,  the
collision term is formulated as
\begin{equation}
C_{1+2\leftrightarrow3+4}=C_{1+2\rightarrow3+4}-C_{3+4\rightarrow1+2}\thinspace,\label{eq:-2}
\end{equation}
with
\begin{equation}
C_{1+2\rightarrow3+4}=\int d\Pi_{1}d\Pi_{2}d\Pi_{3}d\Pi_{4}[f_{1}f_{2}(1\pm f_{3})(1\pm f_{4})]|{\cal M}|^{2}(2\pi)^{4}\delta^{4}\thinspace,\label{eq:-6}
\end{equation}
where $d\Pi_{i}\equiv\frac{d^{3}\mathbf{p}_{i}}{2E_{i}(2\pi)^{3}}$;
  $f_{i}$ is the momentum distribution function of particle $i$;
${\cal M}$ denotes the matrix element of the process (see Tab.~\ref{tab:M2});
 and $\delta^{4}$ denotes the delta function responsible for momentum
conservation. The ``$\pm$'' sign takes ``$+$'' for bosons and
``$-$'' for fermions. 

For one-to-two or two-to-one processes, the collision terms are similar
except that one of the initial or final particles in Eq.~\eqref{eq:-6}
is removed.

\begin{table}[t]
\caption{Matrix elements for relevant processes considered in this work. Two-to-two
processes are expressed in terms of the Mandelstam variables ($s$,
$t$, $u$). The matrix elements are computed assuming either $m_{\nu_{R}}\ll m_{\chi},\ m_{\phi}$
(light-$\nu_{R}$ approx.) or $m_{\nu_{R}}\gg m_{\chi},\ m_{\phi}$
(heavy-$\nu_{R}$ approx., assuming Majorana mass). The mass splitting
parameter $\delta m$ is defined by $\delta m^{2}\equiv m_{\phi}^{2}-m_{\chi}^{2}$.
\label{tab:M2} }

\centering
\begin{tabular}{cccc}
\hline 
Processes & $|{\cal M}|^{2}$ & Processes & $|{\cal M}|^{2}$\tabularnewline
\hline 
\includegraphics[width=3cm]{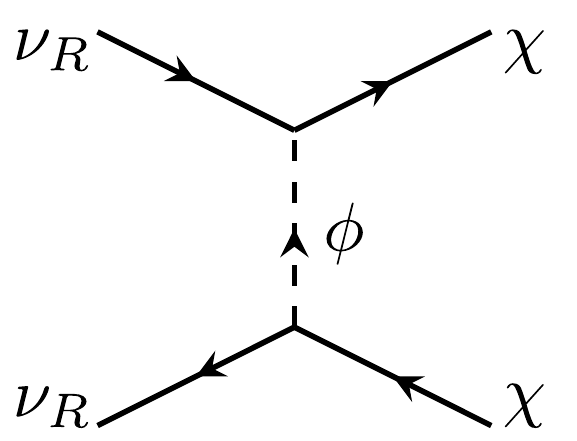} & \raisebox{1.2cm}{\begin{minipage}[t]{0.2\textwidth}$y^{4}\left(\frac{t-m_{\chi}^{2}}{t-m_{\phi}^{2}}\right)^{2}$\\[1mm] 
light-$\nu_{R}$ approx. \end{minipage}} & \includegraphics[width=3cm]{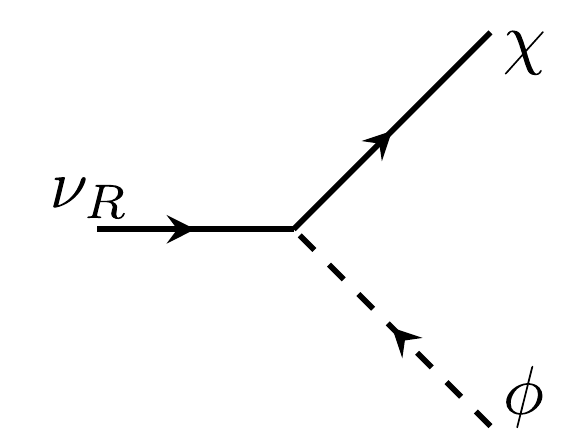} & \raisebox{1.2cm}{\begin{minipage}[t]{0.2\textwidth}$y^{2}m_{\nu_{R}}^{2}$
\\[1mm]  heavy-$\nu_{R}$ approx. \end{minipage}}  \tabularnewline
\includegraphics[width=3cm]{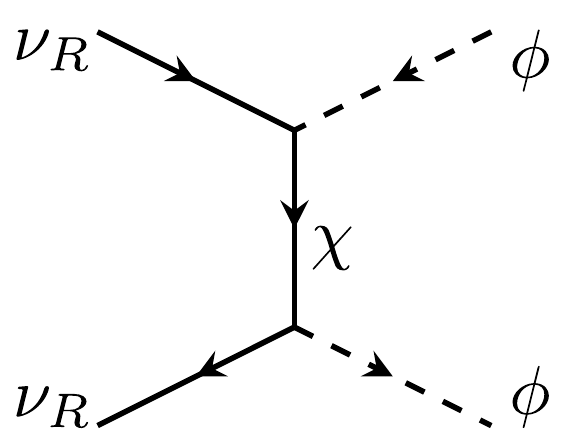} & \raisebox{1.2cm}{\begin{minipage}[t]{0.2\textwidth}$y^{4}\frac{tu-m_{\phi}^{4}}{\left(t-m_{\chi}^{2}\right)^{2}}$
\\[1mm]  light-$\nu_{R}$ approx. \end{minipage}} & \includegraphics[width=3cm]{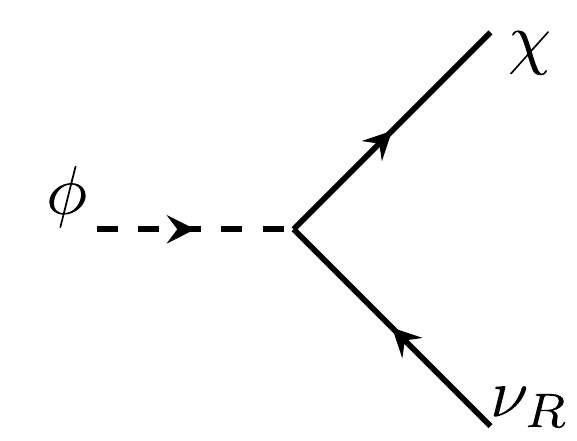} & \raisebox{1.2cm}{\begin{minipage}[t]{0.2\textwidth}$y^{2}\left(m_{\phi}^{2}-m_{\chi}^{2}\right)$
\\[1mm]  light-$\nu_{R}$ approx. \end{minipage}}\tabularnewline
\includegraphics[width=3cm]{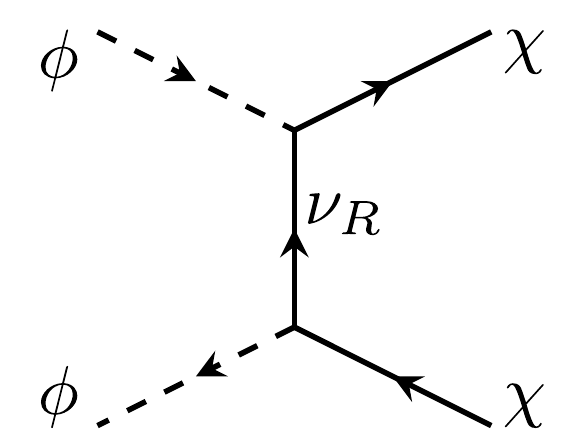} & \raisebox{2cm}{\begin{minipage}[t]{0.2\textwidth}$y^{4}\frac{2tm_{\phi}^{2}-\delta m^{4}-ts-t^{2}}{t^{2}}$
\\[1mm]  light-$\nu_{R}$ approx. \\[1mm]  $y^{4}\frac{s-2m_{\chi}^{2}}{m_{\nu_{R}}^{2}}$
\\[1mm]   heavy-$\nu_{R}$ approx. \end{minipage}} & \includegraphics[width=3cm]{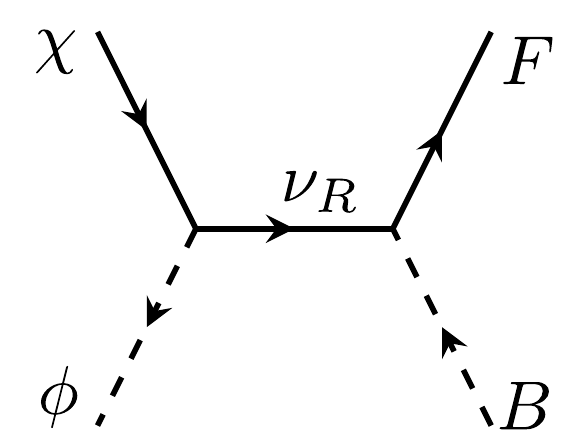} & \raisebox{2.0cm}{\begin{minipage}[t]{0.2\textwidth}$y^{2}y'^{2}\frac{s+t-m_{\phi}^{2}}{s}$
\\[1mm]  light-$\nu_{R}$ approx. \\[1mm]  $y^{2}y'^{2}\frac{m_{\chi}^{2}-t}{m_{\nu_{R}}^{2}}$
\\[1mm]  heavy-$\nu_{R}$ approx. \end{minipage}}\tabularnewline
\hline 
\end{tabular}
\end{table}

Since the SM thermal bath has much more degrees of freedom than the
dark sector, the total energy density of the early universe is dominated
by the SM content.  The Hubble parameter is determined by 
\begin{equation}
H=\sqrt{\frac{8\pi}{3m_{{\rm pl}}^{2}}\left(\rho_{{\rm SM}}+\rho_{{\rm dark}}\right)}=g_{H}\frac{T_{{\rm SM}}^{2}}{m_{{\rm pl}}}\thinspace,\label{eq:-7}
\end{equation}
 where $\rho_{{\rm SM}}$ and $T_{{\rm SM}}$ denote the SM energy
density and temperature, $\rho_{{\rm dark}}$ denotes the energy density
of the $\nu_{R}$-$\chi$-$\phi$ sector, and $m_{{\rm pl}}=1.22\times10^{19}$
GeV. For convenience, we have defined $g_{H}\equiv\sqrt{8\pi^{3}g_{\star}/90}$
where $g_{\star}$ is the effective number of degrees of freedom in
$\rho_{{\rm SM}}$ and $\rho_{{\rm dark}}$. For the SM $g_{\star}$,
we adopt the results in Figure 2.2 of Ref.~\cite{Wallisch:2018rzj}.
For additional species that are produced via freeze-in and stay in
the freeze-in regime, their energy densities are low. For additional
species that undergo non-relativistic freeze-out, their contributions
to $g_{\star}$ are also suppressed at the moment of freeze-out. Therefore,
only for Cases EF and EE with light $\nu_{R}$, $g_{\star}$ receives
a sizable contribution from $\rho_{{\rm dark}}$. 

The SM entropy density, $s_{{\rm SM}}=2\pi^{2}g_{\star}^{(s)}T_{{\rm SM}}^{3}/45$,
depends on a slightly different effective number of degrees of freedom,
$g_{\star}^{(s)}$. In most cases, we neglect the small difference
and assume $g_{\star}^{(s)}\approx g_{\star}$. 

The total entropy of the SM thermal bath in a comoving volume is approximately
conserved, i.e.  $d(s_{{\rm SM}}a^{3})/dt\approx0$ where $a$ is
the scale factor in the FRW metric. Using $H=a^{-1}da/dt$, we obtain
\begin{equation}
\frac{ds_{{\rm SM}}}{dt}+3Hs_{{\rm SM}}\approx0\thinspace.\label{eq:s-con}
\end{equation}

In the freeze-in regime, $n_{X}$ can be computed as follows. Using
Eq.~\eqref{eq:s-con} together with Eq.~\eqref{eq:-1} one gets $dn_{X}/ds_{{\rm SM}}=(C_{X}-3Hn_{X})/(-3Hs_{{\rm SM}})$
which can be written as
\begin{equation}
d\left(\frac{n_{X}}{s_{{\rm SM}}}\right)=-\frac{C_{X}}{3Hs_{{\rm SM}}^{2}}ds_{{\rm SM}}\thinspace,\ \ \text{or}\ \ n_{X}=s_{{\rm SM}}\int_{s_{{\rm SM}}}^{\infty}\frac{C_{X}}{3H\tilde{s}_{{\rm SM}}^{2}}d\tilde{s}_{{\rm SM}}\thinspace,\label{eq:-8}
\end{equation}
where $H$ in the integral depends on the integration variable $\tilde{s}_{{\rm SM}}$.
 If $g_{\star}$ does not change significantly during the relevant
epoch, the integral in Eq.~\eqref{eq:-8} can be written in terms
of $T_{{\rm SM}}$: 
\begin{equation}
n_{X}\approx T_{{\rm SM}}^{3}\frac{m_{{\rm pl}}}{g_{H\text{f.i.}}}\int_{T_{{\rm SM}}}^{\infty}C_{X}\tilde{T}_{{\rm SM}}^{-6}d\tilde{T}_{{\rm SM}}\thinspace,\label{eq:-9}
\end{equation}
where $g_{H{\rm f.i.}}$ denotes the freeze-in value of $g_{H}$,
i.e. it denotes the value of $g_{H}$ at the point when the integrand
peaks. Taking a decay process for example, it should be when the decaying
particle becomes non-relativistic.

After freeze-in, the comoving number density of $X$ is conserved
and $n_{X}/T_{{\rm SM}}^{3}$ would remain as a constant if $T_{{\rm SM}}$
scales as $T_{{\rm SM}}\propto1/a$.   However, due to many subsequent
annihilation of SM species, $T_{{\rm SM}}$ actually scales as $a^{-1}g_{\star}^{-1/3}$.
Therefore, as the universe cools down, the variation of $g_{\star}$
leads to the following correction to $n_{X}$: 

\begin{equation}
n_{X}\to n_{X}\frac{g_{\star}}{g_{\star\text{f.i.}}}\thinspace,\label{eq:-29}
\end{equation}
where $g_{\star\text{f.i.}}$ denotes the freeze-in value of $g_{\star}$,
similar to $g_{H{\rm f.i.}}$ introduced above.

Let us finally comment on the use of the integrated Boltzmann equation
\eqref{eq:-1} and the momentum distribution functions, denoted by
$f_{X}$ for $X=\nu_{R}$, $\chi$,  or $\phi$. For later convenience,
we also denote the momentum distribution function of $X$ in thermal
equilibrium by $f_{X}^{{\rm eq}}$. If $X$ is not in thermal equilibrium,
$f_{X}$ can be very different from $f_{X}^{{\rm eq}}$. 

Strictly speaking, for non-thermal $f_{X}$, one needs to solve the
more fundamental Boltzmann equation for $f_{X}$ rather than the integrated
one for $n_{X}$ but under certain circumstances the latter suffices
for accurate calculations. For non-thermal species being produced
(frozen-in) from a thermal sector, the back-reaction is negligible
and those factors due to Fermi-Dirac or Bose-Einstein statistics {[}such
as $(1\pm f_{3})(1\pm f_{4})$ in Eq.~\eqref{eq:-6}{]} are also negligible.
So the collision term mainly depends on quantities of thermal species
and hence can be well determined. When studying the connection between
two thermal or almost thermal sectors (such as the freeze-out scenario),
the integrated form could also be approximately used because the collision
term can be written in terms of number densities. 

For non-thermal species being produced from another non-thermal species,
however, the spectral shape of the latter can significantly affect
the result. In this case, one needs more elaborated treatments involving
numerical techniques and analytical approximations, as we will elucidate
in Sec.~\ref{subsec:Cases-FF}.

\section{ DM relic abundance \label{sec:DM-relic}}

With the framework set up in Sec.~\ref{sec:Framework}, we now start
to investigate the four generic cases proposed in Fig.~\ref{fig:4-cases}.
The goal is to derive generic formulae for the DM relic abundance
and also the criteria for identifying the four cases. The results
are summarized in Tab.~\ref{tab:result1}.

\begin{table}[h]
\centering

\caption{\label{tab:result1} Results of our analyses for the four generic
cases.   The quantities $R_{1}$ and $R_{2}$, typically ranging
from ${\cal O}(0.1)$ to ${\cal O}(10)$ for GeV-TeV particles, are
defined in Eq.~\eqref{eq:-11} and Eq.~\eqref{eq:-21}. }

\begin{tabular}{c|cc|cc}
\hline 
Cases & \multicolumn{2}{c|}{light $\nu_{R}$ limit ($m_{\chi,\phi}\gg m_{\nu_{R}}$)} & \multicolumn{2}{c}{heavy $\nu_{R}$ limit ($m_{\nu_{R}}\gg m_{\chi,\phi}$)}\tabularnewline
\hline 
 & Valid range of $y$ & $\Omega_{\chi}h^{2}/0.12$ & Valid range of $y$ & $\Omega_{\chi}h^{2}/0.12$\tabularnewline
\cline{2-5} \cline{3-5} \cline{4-5} \cline{5-5} 
\multirow{1}{*}{EE} & $y>2.7\times10^{-4}R_{1}$ & Eq.~\eqref{eq:o-EEFE-1} & $y>1.7\times10^{-7}R_{2}$ & Eq.~\eqref{eq:o-EE-2}\tabularnewline
\multirow{1}{*}{EF} & $y<2.7\times10^{-4}R_{1}$ & Eq.~\eqref{eq:o-EFFF-1} or \eqref{eq:o-EFFF-2}  & $y<1.7\times10^{-7}R_{2}$ & Eq.~\eqref{eq:o-EF-2}\tabularnewline
\multirow{1}{*}{FF} & $y<2.7\times10^{-4}R_{1}$ & Eq.~\eqref{eq:o-FF} & $y<1.7\times10^{-7}R_{2}$ & Eq.~\eqref{eq:o-FFFE-2}\tabularnewline
\multirow{1}{*}{FE} & $y>2.7\times10^{-4}R_{1}$ & Eq.~\eqref{eq:o-EEFE-1} & $y>1.7\times10^{-7}R_{2}$ & Eq.~\eqref{eq:o-FFFE-2}\tabularnewline
\hline 
\end{tabular}
\end{table}

The results crucially depend on whether $\nu_{R}\to\chi\phi$ is kinematically
allowed ($m_{\nu_{R}}>m_{\chi}+m_{\phi}$) or not ($m_{\nu_{R}}<m_{\chi}+m_{\phi}$)
  because the collision term $C_{\nu_{R}\to\chi\phi}$ is much larger,
roughly by a factor of $4\pi^{2}/y^{2}$, than $C_{2\nu_{R}\to2\chi}$
and $C_{2\nu_{R}\to2\phi}$.  For simplicity, we assume $m_{\nu_{R}}\gg m_{\chi,\phi}$
in the decay-allowed case, and $m_{\nu_{R}}\ll m_{\chi,\phi}$ in
the decay-forbidden case. In what follows, they are referred to as
the heavy and light $\nu_{R}$ limit, respectively.  Increasing or
decreasing $m_{\nu_{R}}$ to a level comparable to $m_{\chi,\phi}$
without  crossing the threshold ($m_{\chi}+m_{\phi}$) would lead
to qualitatively similar results,  though the calculations would
be much more complicated.

\subsection{Light $\nu_{R}$ limit \label{subsec:Decay-forbidden}}

In the light $\nu_{R}$ limit, i.e., $m_{\chi,\phi}\gg m_{\nu_{R}}$,
the analyses and results below are almost independent of $m_{\nu_{R}}$,
which in principle  can vary freely from any scales well below $m_{\chi,\phi}$
 down to zero. If sufficiently light, $\nu_{R}$ might contribute
to $N_{{\rm eff}}$ and hence be constrained by the precision measurement
of $N_{{\rm eff}}$. 

\subsubsection{Case EF in the light $\nu_{R}$ limit \label{subsec:Cases-EF-FF}}

Let us start with a sufficiently small $y$ so that the dark sector
is not  thermalized, while the $\nu_{R}$ sector is in thermal equilibrium.
This corresponds to the EF case in Fig.~\ref{fig:4-cases}. The dark
sector particles are produced from $\nu_{R}$ via $2\nu_{R}\to2\chi$
and $2\nu_{R}\to2\phi$. The latter contributes indirectly to the
production of $\chi$ via  $\phi$ decay. Let us first concentrate
on the $2\nu_{R}\to2\chi$ process. The collision term is calculated
in Appendix~\ref{sec:coll} and can be approximately written as 
\begin{equation}
C_{2\nu_{R}\to2\chi}\approx\frac{y^{4}}{128\pi^{5}}\left[m_{\chi}T_{{\rm SM}}K_{1}\left(\frac{m_{\chi}}{T_{{\rm SM}}}\right)\right]^{2},\label{eq:-10}
\end{equation}
 where  $K_{1}$ is the modified Bessel functions of order $1$.
In the derivation of this result,  we have assumed $\delta m^{2}\equiv m_{\phi}^{2}-m_{\chi}^{2}\ll T_{{\rm SM}}^{2}$,
$m_{\chi}\gg(\delta m,m_{\nu_{R}})$, and the Boltzmann statistics.

Given the collision terms, the number densities of $\chi$ and $\phi$
can be computed by solving the Boltzmann equations which in the freeze-in
regime have solutions of the integral forms in Eqs.~\eqref{eq:-8}
and \eqref{eq:-9}.  Substituting Eq.~\eqref{eq:-10} into Eq.~\eqref{eq:-9}
and integrating it down to a temperature well below $m_{\chi}$, we
obtain the number density of $n_{\chi}$ after freeze-in. In addition,
we also take the $g_{\star}$ correction in Eq.~\eqref{eq:-29} into
account. The result reads:
\begin{equation}
n_{\chi}=\frac{3y^{4}m_{\text{pl}}}{2^{12}\pi^{3}g_{H\text{f.i.}}m_{\chi}}\frac{g_{\star}^{(s)}}{g_{\star\text{f.i.}}^{(s)}}T_{{\rm SM}}^{3}\thinspace.\label{eq:-12}
\end{equation}

Apart from the direct production of $\chi$ particles via $2\nu_{R}\to2\chi$,
there is also an indirect production channel via $2\nu_{R}\to2\phi$
followed by $\phi\to\nu_{R}+\chi$ decay. The lifetime of $\phi$
is much shorter than the age of the universe if $y\gtrsim10^{-20}\cdot(m_{\phi}/{\rm GeV})^{-1/2}$.
Therefore, almost all $\phi$ particles produced in the early universe
will eventually decay to $\nu_{R}$ and $\chi$. Since both $C_{2\nu_{R}\to2\phi}$
and $C_{2\nu_{R}\to2\chi}$ are proportional to $y^{4}$, the production
rates of $\phi$ should be comparable to that of $\chi$. Indeed,
according to the calculation in Appendix \ref{sec:coll} {[}see Eqs.~\eqref{eq:-39}
and \eqref{eq:-40}{]}, the number density of $\phi$ after freeze-in
assuming $\phi$ does not decay would be  
\begin{equation}
n_{\phi}\approx1.87n_{\chi}\thinspace.\label{eq:-42}
\end{equation}
Taking $\phi$ decay into account, each $\phi$ particle produces
one $\chi$ particle via $\phi\to\nu_{R}+\chi$. Hence  the number
density of $\chi$ is increased by $n_{\chi}\to2.87n_{\chi}$. 

It is conventional to write the result in terms of $\Omega_{\chi}h^{2}$
where $\Omega_{\chi}=\rho_{\chi}/\rho_{{\rm cri.}}$ is the ratio
of $\rho_{\chi}$ to the critical energy density $\rho_{{\rm cri.}}$
and $h\equiv H_{0}/(100\ \text{km}/\text{sec}/\text{Mpc})$ with $H_{0}$
the Hubble constant today. In terms of $\Omega_{\chi}h^{2}$, the
result reads (including the contribution of $\phi$ decay): 
\begin{equation}
\Omega_{\chi}h^{2}\approx0.12\left(\frac{y}{3.8\times10^{-6}}\right)^{4}\left(\frac{106.75}{g_{\star\text{f.i.}}}\right)^{\frac{3}{2}}\thinspace.\label{eq:o-EFFF-1}
\end{equation}
Note that, as mentioned below Eq.~\eqref{eq:-7}, $g_{\star}$ in
this case receives an addition contribution from light $\nu_{R}$.
So the value of $g_{\star\text{f.i.}}$ should be the SM value at
freeze-in plus $7/8$, assuming a single flavor of $\nu_{R}$.

Eq.~\eqref{eq:o-EFFF-1} implies that to generate $\Omega_{\chi}h^{2}=0.12$
in the EF case, one needs $y\sim10^{-6}$ for thermalized $\nu_{R}$
and $10\lesssim g_{\star\text{f.i.}}\lesssim10^{2}$.  This is consistent
with the result in Ref.~\cite{Hufnagel:2021pso}, see Fig.~3 therein.

The validity of the above analysis for the EF case is based on the
assumption that $y$ is sufficiently small. In this regime, the back-reactions
$2\chi\to2\nu_{R}$  is negligible due to the low density of $\chi$
particles.   Now let us increase $y$, which will eventually lead
to a high back-reaction rate comparable to the production rate. Then
the equilibrium between $\nu_{R}$ and $\chi$ will be established,
leading to $n_{\chi}=n_{\nu_{R}}$.  Substituting $n_{\chi}=n_{\nu_{R}}$
into Eq.~\eqref{eq:-12} and solving  it for $y$, we obtain the following
solution:   
\begin{equation}
y_{{\rm eq}}=2.7\times10^{-4}R_{1},\ \ R_{1}\equiv\left(\frac{g_{H}}{17.2}\frac{m_{\chi}}{{\rm GeV}}\right)^{1/4}.\label{eq:-11}
\end{equation}
  For $y\lesssim y_{{\rm eq}}$, the freeze-in calculation is approximately
 valid. For $y\gtrsim y_{{\rm eq}}$,  it becomes the EE case, which
will be discussed in Sec.~\ref{subsec:Cases-EE-FE-L}.

Finally we would like to comment on a potentially important contribution
from the off-shell $\nu_{R}$ decay. Although $\nu_{R}$ is lighter
than $\chi$ and $\phi$, off-shell $\nu_{R}$ can be produced from
SM particle scattering and then decay to $\chi$ and $\phi$. This
part of contribution depends on the coupling of $\nu_{R}$ to the
SM. In particular, in the presence of a relatively strong $\nu_{R}$-SM
interaction,  the effective thermal mass of $\nu_{R}$ might actually
exceed $m_{\chi}+m_{\phi}$ so that the production rate of the forbidden
channel $\nu_{R}\to\chi\phi$ can be considerably large.    This
is very similar to the plasmon decay ($\gamma^{*}\to2\nu$) used to
constrain neutrino magnetic moments~\cite{Raffelt1996,Capozzi:2020cbu,Li:2022dkc}.
A dedicated treatment requires taking finite-temperature effects into
account consistently in both scattering and decay processes~\cite{Li:2022rde,Li:2023ewv},
which will be studied in our future work. 

Here we provide a straightforward estimate of this contribution assuming
that the finite-temperature effects are negligible and that $\nu_{R}$
couples to a pair of SM particles with a different coupling $y'$.
The collision term for two light SM particles scattering to produce
$\chi$ and $\phi$ is 
\begin{equation}
C_{{\rm SM}\to\nu_{R}^{*}\to\chi\phi}=\frac{y^{2}y'^{2}}{256\pi^{5}}\left[m_{\chi}T_{{\rm SM}}K_{1}\left(\frac{m_{\chi}}{T_{{\rm SM}}}\right)\right]^{2}.\label{eq:-68}
\end{equation}
Comparing it to Eq.~\eqref{eq:-10}, one can see that this channel
dominates when $y'>\sqrt{2}y$. For $y'\gg y$, the DM relic abundance
is given by 
\begin{equation}
\Omega_{\chi}h^{2}\approx0.12\left(\frac{y}{2.5\times10^{-6}}\right)^{2}\left(\frac{y'}{10^{-5}}\right)^{2}\left(\frac{106.75}{g_{\star\text{f.i.}}}\right)^{\frac{3}{2}}\thinspace.\label{eq:o-EFFF-2}
\end{equation}
For $y'\ll y$, one should use Eq.~\eqref{eq:o-EFFF-1} instead.

\subsubsection{Case FF in the light $\nu_{R}$ limit \label{subsec:Cases-FF}}

\begin{figure}
\centering

\includegraphics[width=0.6\textwidth]{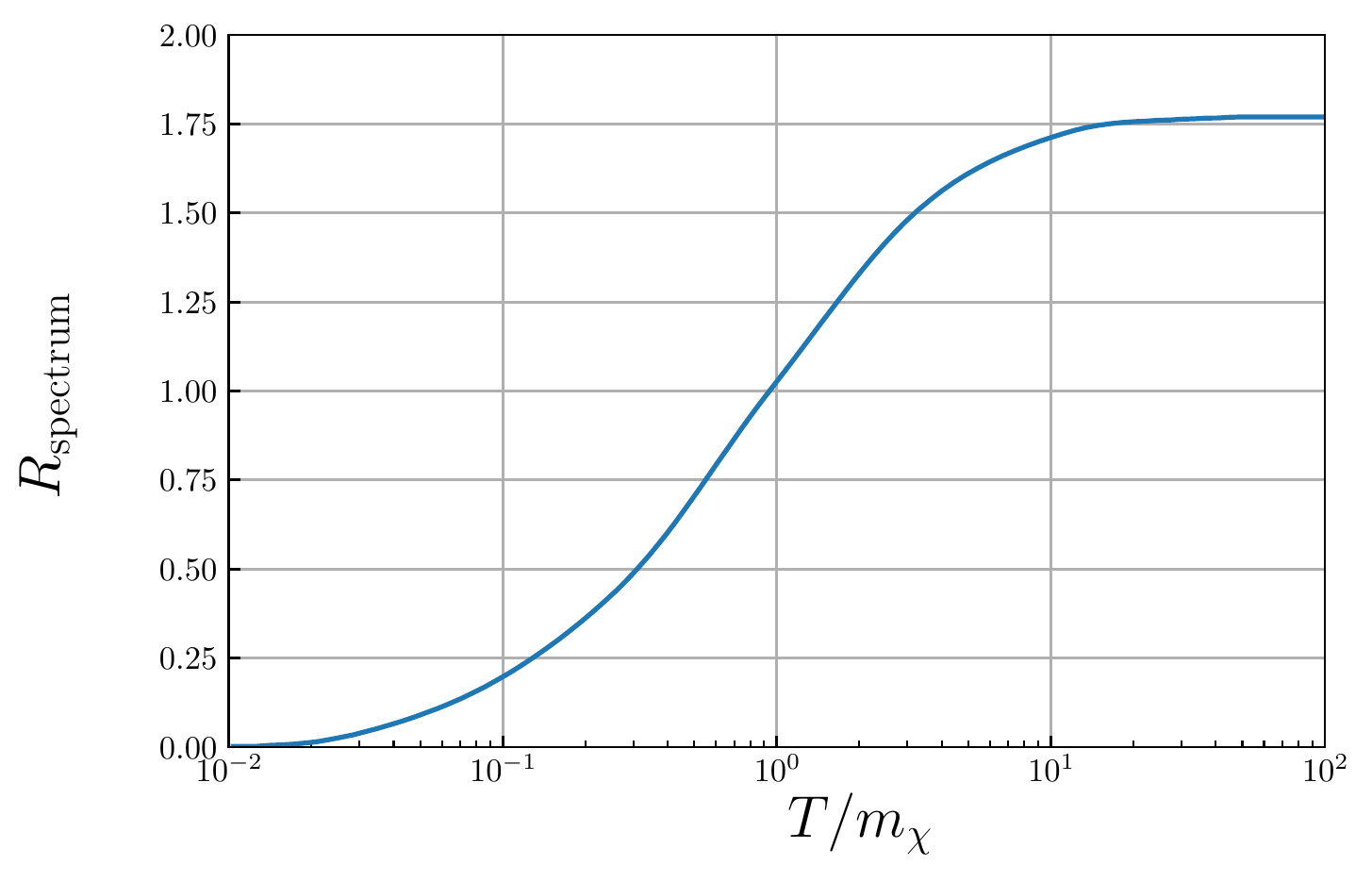}

\caption{The spectral distortion factor $R_{\text{spectrum}}$ defined in Eq.~\eqref{eq:-85}
as a function of $T$. \label{fig:R-spect} }

\end{figure}

In the FF case, $\nu_{R}$ is not in thermal equilibrium so the result
depends on the specific form of the momentum distribution function
of $\nu_{R}$. For the FF case, we modify  Eq.~\eqref{eq:-10} as
follows:
\begin{equation}
C_{2\nu_{R}\to2\chi}\approx\frac{y^{4}}{128\pi^{5}}\left\langle \frac{f_{\nu_{R}}}{f_{\nu_{R}}^{{\rm eq}}}\right\rangle ^{2}\left[m_{\chi}T_{{\rm SM}}K_{1}\left(\frac{m_{\chi}}{T_{{\rm SM}}}\right)\right]^{2},\label{eq:-84}
\end{equation}
where $\left\langle f_{\nu_{R}}/f_{\nu_{R}}^{{\rm eq}}\right\rangle $
is, by definition, to absorb the difference between FF and EF collisions
terms for $2\nu_{R}\to2\chi$. Since $\left(n_{\nu_{R}},\ n_{\nu_{R}}^{{\rm eq}}\right)=\int\left(f_{\nu_{R}},\ f_{\nu_{R}}^{{\rm eq}}\right)\frac{d^{3}p}{(2\pi)^{3}}$,
we expect $\left\langle f_{\nu_{R}}/f_{\nu_{R}}^{{\rm eq}}\right\rangle $
to be around the same order of magnitude as $n_{\nu_{R}}/n_{\nu_{R}}^{{\rm eq}}$.
Hence we introduce a spectral distortion factor defined as
\begin{equation}
R_{\text{spectrum}}\equiv\left\langle \frac{f_{\nu_{R}}}{f_{\nu_{R}}^{{\rm eq}}}\right\rangle \left(\frac{n_{\nu_{R}}}{n_{\nu_{R}}^{{\rm eq}}}\right)^{-1}.\label{eq:-85}
\end{equation}
In the absence of spectral distortion ($f_{\nu_{R}}\propto f_{\nu_{R}}^{{\rm eq}}$),
we have $R_{\text{spectrum}}=1$. For light $\nu_{R}$ produced from
heavy particle decay, the spectral shape of $f_{\nu_{R}}$ in the
freeze-in regime can be computed analytically~\cite{Heeck:2017xbu,Decant:2021mhj}---see
also Appendix~\ref{sec:app-f} for a brief review. With the analytical
expression for $f_{\nu_{R}}$, we can compute $C_{2\nu_{R}\to2\chi}$
numerically\footnote{Here we have used the  code available at \url{https://github.com/xunjiexu/Thermal_Monte_Carlo}.},
extract $\left\langle f_{\nu_{R}}/f_{\nu_{R}}^{{\rm eq}}\right\rangle $
in Eq.~\eqref{eq:-84}, and then obtain the spectral distortion factor
$R_{\text{spectrum}}$, which is presented in Fig.~\ref{fig:R-spect}. 

Using the numerical value of $R_{\text{spectrum}}$, we repeat the
analysis in Sec.~\ref{subsec:Cases-EF-FF} and find that the non-thermal
$f_{\nu_{R}}$ results in the following correction to the DM relic
abundance in Eq.~\eqref{eq:o-EFFF-1}:
\begin{equation}
\Omega_{\chi}h^{2}\to1.74\left(\frac{n_{\nu_{R}}}{n_{\nu_{R}}^{{\rm eq}}}\right)^{2}\Omega_{\chi}h^{2}\thinspace,\label{eq:-86}
\end{equation}
Therefore, for the FF case, the DM relic abundance abundance can be
written as
\begin{equation}
\Omega_{\chi}h^{2}\approx0.12\left(\frac{n_{\nu_{R}}}{n_{\nu_{R}}^{{\rm eq}}}\right)^{2}\left(\frac{y}{3.3\times10^{-6}}\right)^{4}\left(\frac{106.75}{g_{\star\text{f.i.}}}\right)^{\frac{3}{2}}\thinspace.\label{eq:o-FF}
\end{equation}
Here we have assumed that $n_{\nu_{R}}/n_{\nu_{R}}^{{\rm eq}}$ is
approximately a  constant during the epoch of the second freeze-in
of FF. This is a good approximation if the two freeze-in processes
of FF occur at well separated temperature scales. If the two freeze-in
temperatures are close to each other, then Eq.~\eqref{eq:o-FF} is
not valid but it can be used as a crude estimate. The exact result
in this case can be obtained by numerical integration. 

\subsubsection{Cases EE and FE in the light $\nu_{R}$ limit \label{subsec:Cases-EE-FE-L}}

The EE and FE cases in Fig.~\ref{fig:4-cases} require that the dark
sector keeps thermal equilibrium with $\nu_{R}$.  The criteria for
such cases are just mentioned above. 

In these two cases, we have $n_{\chi}\approx n_{\nu_{R}}$ at temperatures
well above both masses. As the temperature falls below $\sim m_{\chi}$,
a large amount of  $\chi$ particles will annihilate to $\nu_{R}$,
leaving only a small portion of $\chi$ particles due to the well-known
freeze-out mechanism. Since the freeze-out mechanism has been extensively
studied, we refer to textbooks~\cite{Kolb:1990vq,Dodelson2003} and
lectures~\cite{Plehn:2017fdg,Lisanti:2016jxe,Lin:2019uvt} for a
full calculation starting from the Boltzmann equation. Here one can
adapt the known result of  the standard freeze-out calculation to
our framework, with the key difference being that  the $\nu_{R}$-dark
sector may  have a different temperature compared to the SM one. The
details are presented in Appendix~\ref{sec:freeze-out} and the results
are summarized below. 

For convenience, we define two ratios
\begin{equation}
\epsilon\equiv T_{\chi}/T_{{\rm SM}}\thinspace,\ x\equiv m_{\chi}/T_{\chi}\thinspace,\label{eq:x}
\end{equation}
and denote their corresponding freeze-out values by $\epsilon_{{\rm f.o.}}$
and $x_{{\rm f.o.}}$. In the EE case, we have $\epsilon_{{\rm f.o.}}=1$.
In the FE case, $\epsilon_{{\rm f.o.}}$ depends on the abundance
of $\nu_{R}$ produced from the SM sector. The value of $x_{{\rm f.o.}}$
according to Appendix~\ref{sec:freeze-out} is approximately given
by
\begin{equation}
x_{{\rm f.o}}=35.5+9.5\log_{10}\left[y\cdot\epsilon_{{\rm f.o}}^{1/2}\cdot\left(\frac{m_{\chi}}{{\rm GeV}}\frac{g_{H}}{17.2}\right)^{-1/4}\right].\label{eq:-14}
\end{equation}
In the standard WIMP paradigm, $x_{{\rm f.o.}}$ typically varies
around $20\sim25$.

   The relic abundance of DM is given by
\begin{equation}
\Omega_{\chi}h^{2}=0.12\frac{x_{{\rm f.o.}}\epsilon_{{\rm f.o.}}}{\sqrt{g_{{\rm \star f.o.}}}}\cdot\frac{1.4\times10^{-9}\ \text{GeV}^{-2}}{\langle\sigma v\rangle}\thinspace,\label{eq:-17}
\end{equation}
where $\langle\sigma v\rangle$ is the velocity-averaged cross section,
widely used in the WIMP paradigm. It is related to the collision term
by
\begin{equation}
\langle\sigma v\rangle\equiv\frac{1}{n_{\chi}^{2}}C_{2\chi\to2\nu_{R}}\thinspace,\label{eq:-15}
\end{equation}
and in the non-relativistic regime given by
\begin{equation}
\langle\sigma v\rangle\approx\frac{y^{4}}{32\pi m_{\chi}^{2}}\thinspace.\label{eq:-43}
\end{equation}
Substituting Eq.~\eqref{eq:-43} into Eq.~\eqref{eq:-17}, we obtain
\begin{equation}
\Omega_{\chi}h^{2}=0.12\left(\frac{m_{\chi}}{100\ {\rm GeV}}\right)^{2}\left(\frac{0.19}{y}\right)^{4}\frac{x_{{\rm f.o.}}\epsilon_{{\rm f.o.}}}{\sqrt{g_{{\rm \star f.o.}}}}\thinspace.\label{eq:o-EEFE-1}
\end{equation}
Similar to the discussion below Eq.~\eqref{eq:o-EFFF-1}, $g_{\star}$
in this case receives an addition contribution from light $\nu_{R}$.
So the value of $g_{\star\text{f.o.}}$ should be the SM value at
freeze-out plus $7\epsilon_{{\rm f.o.}}^{4}/8$, assuming a single
flavor of $\nu_{R}$. 

Note that in the EE and FE cases, the influence of $\phi\to\nu_{R}\chi$
on the DM abundance should be small if $\delta m^{2}x_{{\rm f.o.}}>2m_{\chi}^{2}$
because $n_{\phi}/n_{\chi}\approx\exp\left[\left(m_{\chi}-m_{\phi}\right)/T\right]$
is small at the freeze-out temperature. Taking $x_{{\rm f.o.}}=25$
and $\delta m^{2}=0.2m_{\chi}^{2}$ for example, we get $n_{\phi}/n_{\chi}=9.2\%$.
Although the observed value of $\Omega_{\chi}h^{2}$ has been determined
very precisely ($\sim1\%$), a 9\% variation of $\Omega_{\chi}h^{2}$
can be easily absorbed into a small variation of $y$, which is a
rather free parameter in the model. So here we neglect the contribution
of $\phi\to\chi\nu_{R}$ to the DM abundance.

A particularly interesting feature of the FE case is that due to the
DM annihilation at a relatively late epoch, the abundance of $\nu_{R}$
might be substantially enhanced, causing observable changes to $N_{{\rm eff}}$
if $\nu_{R}$ are sufficiently light.

\subsection{Heavy $\nu_{R}$ limit }

In the heavy $\nu_{R}$ limit ($m_{\nu_{R}}\gg m_{\chi,\phi}$), $\nu_{R}\to\chi\phi$
as well as its back-reaction $\chi\phi\to\nu_{R}$ is kinematically
allowed.  In the presence of $\nu_{R}\leftrightarrow\chi\phi$,
the energy conversion  between the two sectors via such two-to-one
or one--to-two processes is much more efficient than that via two-to-two
processes considered in Sec.~\ref{subsec:Decay-forbidden}. Therefore,
the contributions of $2\nu_{R}\leftrightarrow2\chi$ and $2\nu_{R}\leftrightarrow2\phi$
will be neglected in what follows.

\subsubsection{Case EF in the heavy $\nu_{R}$ limit\label{subsec:Case-EF-H}}

Similar to the analysis in Sec.~\ref{subsec:Decay-forbidden}, we
start with a sufficiently small $y$ so that the dark sector is not
thermalized. On the other hand, the $\nu_{R}$ sector may or may not
be in thermal equilibrium with the SM thermal bath, depending on the
interaction of $\nu_{R}$ with the SM. Let us first consider that
the $\nu_{R}$-SM coupling is sufficiently large so that the $\nu_{R}$
sector has been thermalized.   This leads to the EF case.   The
collision term for the dominant DM production process $\nu_{R}\to\chi\phi$
reads
\begin{equation}
C_{\nu_{R}\to\chi\phi}=y^{2}\frac{m_{\nu_{R}}^{3}T_{{\rm SM}}}{32\pi^{3}}K_{1}\left(\frac{m_{\nu_{R}}}{T_{{\rm SM}}}\right).\label{eq:-18}
\end{equation}

Now substituting Eq.~\eqref{eq:-18} into Eq.~\eqref{eq:-9} and taking
the $g_{\star}$ correction in Eq.~\eqref{eq:-29} into account, 
we obtain  the values of $n_{\chi}$ and $n_{\phi}$ after freeze-in:
\begin{equation}
n_{\chi}=n_{\phi}=\frac{3y^{2}m_{\text{pl}}}{64\pi^{2}m_{\nu_{R}}g_{H\text{f.i.}}}\frac{g_{\star}^{(s)}}{g_{\star\text{f.i.}}^{(s)}}T_{{\rm SM}}^{3}\thinspace.\label{eq:-19}
\end{equation}

As we have assumed $m_{\nu_{R}}>m_{\phi}+m_{\chi}$,  $\phi$ cannot
decay to $\chi$ and $\nu_{R}$. However, this does not mean that
$\phi$ would be stable. As long as the $\nu_{R}$-SM coupling is
not extremely suppressed, $\phi$ can eventually decay to $\chi$
and some SM final states with $\nu_{R}$ as an intermediate state.
And the lifetime of $\phi$  is generally expected to be much shorter
than the age of the universe. So for today's $n_{\chi}$, we should
take $n_{\chi}\to n_{\chi}+n_{\phi}=2n_{\chi}$. Consequently, today's
DM relic abundance is given by
\begin{equation}
\Omega_{\chi}h^{2}=0.12\left(\frac{y}{1.25\times10^{-11}}\right)^{2}\left(\frac{m_{\chi}/m_{\nu_{R}}}{0.01}\right)\left(\frac{106.75}{g_{\star\text{f.i.}}}\right)^{\frac{3}{2}}\thinspace.\label{eq:o-EF-2}
\end{equation}

Note that there is an interesting exception known as the SuperWIMP
mechanism~\cite{Covi:1999ty,Feng:2003xh,Feng:2003uy} in which $\nu_{R}$
had been in thermal equilibrium but froze out before the freeze-in
process $\nu_{R}\to\chi\phi$ started. Strictly speaking, this does
not belong to the EF case because $\nu_{R}$ has already decoupled
from the thermal bath when the majority of DM is being produced via
$\nu_{R}\to\chi\phi$. We will discuss this exception in Sec.~\ref{subsec:Cases-FF-FE-H}.

Similar to Eq.~\eqref{eq:-11}, there is also an upper limit of $y$
for the validity of the freeze-in calculation. This is determined
by equating $n_{\chi}$ to its equilibrium value. By solving it for
$y$, we obtain the equilibrium criterion:
\begin{equation}
y_{{\rm eq}}=1.7\times10^{-7}R_{2},\ R_{2}\equiv\left(\frac{m_{\nu_{R}}}{1\ \text{TeV}}\cdot\frac{g_{H}}{17.2}\right)^{1/2}.\label{eq:-21}
\end{equation}
For $y\lesssim y_{{\rm eq}}$, the above calculation for the EF case
is valid. Otherwise, it becomes the EE case, as we will discuss below.

\subsubsection{Case EE in the heavy $\nu_{R}$ limit\label{subsec:Case-EE-H}}

For $y$ exceeding the limit in Eq.~\eqref{eq:-21}, $\chi$ and $\phi$
are in thermal equilibrium with the SM thermal bath, entering the
EE case. As the universe cools down to a temperature below $m_{\chi}$
and $m_{\phi}$, the energy and entropy stored in the $\phi$-$\chi$
sector will be released via off-shell $\nu_{R}$ (virtual) to the
SM thermal bath. The calculation for this scenario  requires assumptions
on the $\nu_{R}$-SM interaction. To proceed, we assume that $\nu_{R}$
is coupled to a pair of SM particles ($F$ and $B$) similar to Eq.~\eqref{eq:Yukawa}
with a different coupling $y'$:
\begin{equation}
{\cal L}\supset y'B\nu_{R}F\thinspace.\label{eq:-46}
\end{equation}
Here $F$ is a chiral fermion and $B$ a scalar boson, with masses
well below $m_{\nu_{R}}$.   With the interaction in Eq.~\eqref{eq:-46},
the most efficient channel converting particles in the dark sector
particles to SM species is co-annihilation---see e.g.~\cite{Griest:1990kh,Edsjo:1997bg,Berlin:2017ife}.
Through the $s$-channel tree-level process $\chi\phi\to FB$, each
pair of $\chi$ and $\phi$ can be efficiently converted to a pair
of $F$ and $B$ when the dark sector becomes non-relativistic. The
process stops when $\chi$ and $\phi$ particles are too rare to meet
each other, i.e., when $n_{\chi}=n_{\phi}\sim H/\langle\sigma v\rangle$.
Hence similar to the standard WIMP freeze-out, there is also freeze-out
in the co-annihilation scenario.

The co-annihilation collision term in the non-relativistic regime
 and in the small mass splitting limit reads:
\begin{equation}
C_{\chi\phi\to FB}\approx\frac{m_{\chi}^{3}T^{3}y^{2}y'^{2}}{128\pi^{4}m_{\nu_{R}}^{2}}e^{-2m_{\chi}/T}\thinspace.\label{eq:-47}
\end{equation}
Correspondingly, the velocity-averaged cross section is
\begin{equation}
\langle\sigma v\rangle=\frac{1}{n_{\chi}^{2}}C_{\chi\phi\to FB}\approx\frac{y^{2}y'^{2}}{32x}\left[\frac{e^{-x}}{m_{\nu_{R}}K_{2}(x)}\right]^{2}\approx\frac{y^{2}y'^{2}}{16\pi m_{\nu_{R}}^{2}}.\label{eq:-22-1}
\end{equation}

The freeze-out temperature can be determined by solving $n_{\chi}\langle\sigma v\rangle=H$.
Similar to Eq.~\eqref{eq:-14}, here we also have an approximate solution
(see Appendix~\ref{sec:freeze-out}):
\begin{equation}
x_{{\rm f.o}}\approx22.2+4.88\log_{10}\left[yy'\cdot\left(\frac{m_{\chi}}{\text{GeV}}\right)^{1/2}\cdot\left(\frac{m_{\nu_{R}}}{1\ {\rm TeV}}\right)^{-1}\cdot\left(\frac{g_{H}}{17.2}\right)^{-1/2}\right].\label{eq:-14-2}
\end{equation}

The DM relic abundance can be computed using Eq.~\eqref{eq:-17} with
$x_{{\rm f.o}}$ from Eq.~\eqref{eq:-14-2} and   $\langle\sigma v\rangle$
from Eq.~\eqref{eq:-22-1}. The result reads 
\begin{equation}
\Omega_{\chi}h^{2}=0.12\left(\frac{m_{\nu_{R}}}{1\ {\rm TeV}}\right)^{2}\cdot\left(\frac{0.39}{yy'}\right)^{2}\cdot\left(\frac{x_{{\rm f.o}}}{22.2}\right)\cdot\left(\frac{106.75}{g_{{\rm \star f.o.}}}\right)^{\frac{1}{2}}\thinspace.\label{eq:o-EE-2}
\end{equation}

\subsubsection{Cases FF and FE  in the heavy $\nu_{R}$ limit\label{subsec:Cases-FF-FE-H}}

In these two cases, after a certain amount of $\nu_{R}$ particles
are produced via freeze-in, they will eventually decay to $\phi$
and $\chi$ particles, if we assume that $\nu_{R}\to\chi\phi$ is
the dominant decay mode of $\nu_{R}$. This is expected in the FE
case because $\nu_{R}$ should be more tightly coupled to the dark
sector than to the SM sector. As for the FF case, a considerably large
part of the parameter space should meet the assumption.    Hence
for both cases under this assumption, the  number density of $\chi$
is determined by the number of  $\nu_{R}$ accumulated during the
first freeze-in:    
\begin{equation}
n_{\chi}=2\frac{g_{\star}}{g_{\star\text{f.i}}}T_{{\rm SM}}^{3}\frac{m_{{\rm pl}}}{g_{H\text{f.i}}}\int_{T_{{\rm SM}}}^{\infty}C_{{\rm SM}\to\nu_{R}}\tilde{T}_{{\rm SM}}^{-6}d\tilde{T}_{{\rm SM}}\thinspace.\label{eq:-44}
\end{equation}
Here we have added a factor of $2$ to account for the contribution
of $\phi$ decay---see the discussion below Eq.~\eqref{eq:-19}.

Assuming  $\nu_{R}$ is produced via Eq.~\eqref{eq:-46},  the integral
in Eq.~\eqref{eq:-44} gives
\begin{equation}
\Omega_{\chi}h^{2}=0.12\left(\frac{m_{\chi}/m_{\nu_{R}}}{0.01}\right)\cdot\left(\frac{y'}{1.25\times10^{-11}}\right)^{2}\cdot\left(\frac{106.75}{g_{{\rm \star f.i.}}}\right)^{\frac{3}{2}}\thinspace.\label{eq:o-FFFE-2}
\end{equation}
Note that the result is independent of $y$. Although $y$ can be
used to tell whether it belongs to the FF or FE case, the difference
between the two cases becomes unimportant as long as $\nu_{R}$ dominantly
decays to $\chi$ and $\phi$. The dominance can be guaranteed if
$y\gg y'$ which is plausible if one compares the benchmark value
of $y'=1.25\times10^{-11}$ in Eq.~\eqref{eq:o-FFFE-2} to  $1.7\times10^{-7}$
in Eq.~\eqref{eq:-21}. 

As previously mentioned in Sec.~\ref{subsec:Case-EF-H}, heavy $\nu_{R}$
could allow for the SuperWIMP mechanism~\cite{Covi:1999ty,Feng:2003xh,Feng:2003uy}
in which DM is produced from frozen-out $\nu_{R}$. This scenario
is technically similar to the FF case because $\nu_{R}$ has decoupled
from the SM thermal bath when it becomes responsible for DM production.
 Assuming that $\nu_{R}$ dominantly decays to the dark sector, the
abundance of DM can be estimated from the number density of $\nu_{R}$
after the freeze-out. In the SuperWIMP mechanism, the dominance could
be achieved by setting $m_{B}\geq m_{\nu_{R}}$ or $y\gg y'$. Under
this assumption, the DM relic abundance is given by
\begin{equation}
\Omega_{\chi}h^{2}=0.12\frac{m_{\nu_{R}}}{10\ {\rm GeV}}\frac{m_{\chi}}{50\ {\rm MeV}}\left(\frac{0.019}{y'}\right)^{4}\frac{x_{{\rm f.o.}}}{\sqrt{g_{{\rm \star f.o.}}}}\thinspace.\label{eq:SuperWIMP}
\end{equation}

Here we would like to compare  the FF and FE cases in the heavy $\nu_{R}$
limit with the corresponding ones in the light $\nu_{R}$ limit. In
the FF case with $m_{\nu_{R}}\ll m_{\chi}$, the second freeze-in
($2\nu_{R}\to2\chi$) stops when the $\nu_{R}$ kinetic energy drops
below the mass of $\chi$. In particular, after this freeze-in stops,
a large amount of $\nu_{R}$ may still be present.    In the FF
case with $m_{\nu_{R}}\gg m_{\chi}$, the second freeze-in can only
be stopped by complete depletion of $\nu_{R}$. Hence the production
rate $C_{\nu_{R}\to\chi\phi}$ does not need to compete with the Hubble
expansion, rendering $y$ unimportant as we have just mentioned. 

As for the FE case, the $\nu_{R}$-$\chi$-$\phi$ coupled sector
acquires a certain amount of energy and entropy injected from the
SM thermal bath.  The key difference between $m_{\nu_{R}}\ll m_{\chi}$
and $m_{\nu_{R}}\gg m_{\chi}$ is that the majority of this portion
of energy and entropy will eventually be stored in the lightest species,
which is $\nu_{R}$ (for $m_{\nu_{R}}\ll m_{\chi}$) or $\chi$ (for
$m_{\nu_{R}}\ll m_{\chi}$).  

Finally, we would like to comment on a potentially important two-to-two
process  in the heavy $\nu_{R}$ limit. For heavy $\nu_{R}$, DM
could be directly produced by the scattering of two SM particles,
through an off-shell $\nu_{R}$ as the intermediate state, to $\phi$
and $\chi$. At $T_{{\rm SM}}\ll m_{\nu_{R}}$, this is the dominant
channel for DM production. The collision term of such a process assuming
Eq.~\eqref{eq:-4} is proportional to $y^{2}y'^{2}/m_{\nu_{R}}^{2}$,
suppressed by $m_{\nu_{R}}^{-2}$ in the heavy $\nu_{R}$ limit, whereas
the production via on-shell $\nu_{R}$ decay is exponentially suppressed.
Despite the dominance of the two-to-two process at $T_{{\rm SM}}\ll m_{\nu_{R}}$,
 the overall contribution after integrating over $T_{{\rm SM}}$
is still subdominant compare to that of on-shell $\nu_{R}$ decay.

\subsection{Discussions}

The calculations presented above are all based on the assumption that
the dark sector only interacts with the SM via $\nu_{R}$. However,
in this framework, the singlet scalar $\phi$ can be readily coupled
to the SM Higgs via the Higgs portal. Assuming that the Higgs portal
interaction is ${\cal L}\supset\frac{1}{4}\lambda|H|^{2}|\phi|^{2}$,
to avoid a significant amount of dark sector particles being produced
via the Higgs portal, the coupling $\lambda$ needs to sufficiently
small. 

Here let us briefly estimate the production rate of $\phi$ via $2H\to2\phi$
in the early universe. The matrix element for this process is simply
an energy-independent constant, $|{\cal M}|\sim\lambda$. For $m_{\phi}\gg m_{H}$,
we obtain the following collision term:
\begin{equation}
C_{2H\to2\phi}=\frac{\lambda^{2}}{128\pi^{5}}\left[m_{\phi}T_{{\rm SM}}K_{1}\left(\frac{m_{\phi}}{T_{{\rm SM}}}\right)\right]^{2},\label{eq:-69}
\end{equation}
 which is similar to the $2\nu_{R}\to2\chi$ collision term. So it
implies that for a sufficiently small $\lambda$,  it should be in
the freeze-in regime. Following the usual freeze-in calculation, we
obtain that the abundance of $\phi$ (in the absence of $\phi$ decay)
would be comparable to $\Omega_{\chi}h^{2}=0.12$ if $\lambda\approx2.5\times10^{-11}$.
Therefore, we conclude that for $\lambda$ well below $10^{-11}$,
the Higgs portal interaction can be neglected.

\section{Application: Type-I seesaw dark matter\label{sec:seesaw}}

In this section, we apply our framework to the type-I seesaw model,
i.e., we take the type-I seesaw model and  extend it by a pair of
 dark particles, $\chi$ and $\phi$, with the interaction given by
Eq.~\eqref{eq:Yukawa}. Such an extension can accommodate an absolutely
stable DM candidate and meanwhile still retain the virtue of being
responsible for light neutrino masses via the type-I seesaw mechanism.

The type-I seesaw Lagrangian reads:
\begin{equation}
{\cal L}\supset y_{\nu}\tilde{H}^{\dagger}L\nu_{R}+\frac{1}{2}m_{\nu_{R}}\nu_{R}\nu_{R}+{\rm h.c.}\thinspace,\label{eq:-55}
\end{equation}
where $H=\frac{1}{\sqrt{2}}(0,h+v)^{T}$ is the SM Higgs doublet in
the unitarity gauge,  $\tilde{H}\equiv i\sigma_{2}H^{*}$, and $L=(\nu_{L},\ e_{L})^{T}$
is a lepton doublet. After the electroweak symmetry breaking,  
the above Yukawa interaction gives rises to the Dirac mass term, ${\cal L}\supset m_{D}\nu_{L}\nu_{R}$
with $m_{D}=y_{\nu}v/\sqrt{2}$. Then at low energy scales, $\nu_{L}$
acquires an effective Majorana mass via the seesaw mechanism:
\begin{equation}
m_{\nu_{L}}=m_{D}^{2}/m_{\nu_{R}}\thinspace.\label{eq:-54}
\end{equation}
For simplicity, we ignore the flavor structure and regard $m_{\nu_{L}}$,
$m_{D}$, and  $m_{\nu_{R}}$ all as single-value quantities rather
than $3\times3$ matrices. 

It is noteworthy that the type-I seesaw model itself in principle
allows for one of the $\nu_{R}$'s to be a DM candidate in the keV
regime---see e.g. the so-called $\nu$MSM~\cite{Asaka:2005pn,Asaka:2005an}.
However, since the simplest scenario where the keV sterile neutrino
DM is produced by the Dodelson--Widrow mechanism~\cite{Dodelson:1993je}
has been excluded by $X$-ray and Lyman-$\alpha$ observations, 
 we do not consider the possibility that $\nu_{R}$ servers as a
DM candidate in this work. 

Given the approximately known scale of $m_{\nu_{L}}$, we make use
of the seesaw mass relation \eqref{eq:-54} to determine the Yukawa
coupling:
\begin{align}
y_{\nu}=\frac{1}{v}\sqrt{2m_{\nu_{L}}m_{\nu_{R}}}\approx & 0.057\cdot\left(\frac{m_{\nu_{R}}}{10^{12}\ {\rm GeV}}\right)^{1/2}\label{eq:ya}\\
\approx & 5.7\times10^{-8}\cdot\left(\frac{m_{\nu_{R}}}{1\ {\rm GeV}}\right)^{1/2}\thinspace,\label{eq:yb}
\end{align}
where we have assumed $m_{\nu_{L}}\approx0.1$ eV.  

The seesaw mechanism also leads to the active-sterile neutrino mixing
with the mixing angle given by
\begin{equation}
\theta\approx\sqrt{\frac{m_{\nu_{L}}}{m_{\nu_{R}}}}\approx5.7\times10^{-13}y_{\nu}^{-1}\thinspace.\label{eq:-45}
\end{equation}

Below we will show  that $y_{\nu}$ and $\theta$ determined by the
type-I seesaw relation are sufficiently large to thermalize $\nu_{R}$.
Hence the type-I  seesaw DM is always in the EE or EF case.

\subsection{Production rates of $\nu_{R}$ via Yukawa and gauge interactions
\label{subsec:Yukawa}}

\subsubsection{Above the electroweak scale}

For $m_{\nu_{R}}$ well above the electroweak scale, the dominant
process for $\nu_{R}$ production is $h+\nu_{L}\to\nu_{R}$. The collision
term is
\begin{equation}
C_{h+\nu_{L}\to\nu_{R}}\approx y_{\nu}^{2}\frac{m_{\nu_{R}}^{3}T_{{\rm SM}}}{32\pi^{3}}K_{1}\left(\frac{m_{\nu_{R}}}{T_{{\rm SM}}}\right).\label{eq:-49}
\end{equation}
Using $C_{h+\nu_{L}\to\nu_{R}}$, we can estimate {[}similar to the
derivation of Eq.~\eqref{eq:-21}{]} the criterion for $\nu_{R}$
reaching thermal equilibrium:
\begin{equation}
y_{\nu}\gtrsim1.7\times10^{-7}\cdot\left(\frac{m_{\nu_{R}}}{1\ \text{TeV}}\right)^{1/2},\label{eq:-21-1}
\end{equation}
which is always satisfied according to Eq.~\eqref{eq:ya}. Therefore,
$\nu_{R}$ must have been in thermal equilibrium if $y_{\nu}$ is
determined by the seesaw relation and $m_{\nu_{R}}$ is well above
the electroweak scale. 

\subsubsection{Below the electroweak scale}

For $m_{\nu_{R}}$ well below the electroweak scale, $h+\nu_{L}\to\nu_{R}$
is kinematically forbidden but $h\to\nu_{L}+\nu_{R}$ is allowed.
The collision term is similar:
\begin{equation}
C_{h\to\nu_{L}+\nu_{R}}\approx y_{\nu}^{2}\frac{m_{h}^{3}T_{{\rm SM}}}{32\pi^{3}}K_{1}\left(\frac{m_{h}}{T_{{\rm SM}}}\right),\label{eq:-52}
\end{equation}
where $m_{h}\approx125$ GeV is the Higgs mass. 

In addition to the Higgs decay, $\nu_{R}$ can also be produced via
$Z$ and $W^{\pm}$ decays. One can compute the decay widths either
in the mass-eigenstate basis or in the chiral basis. The results from
the two different approaches are the same. Taking $Z$ decay as an
example, in the mass-eigenstate basis, it is straightforward to obtain
$C_{Z\to\nu_{L}+\nu_{R}}\approx\theta^{2}C_{Z\to\nu_{L}+\nu_{L}}$,
though strictly speaking one should replace $\nu_{L}$ and $\nu_{R}$
with their respective dominant mass eigenstates.  In the chiral basis,
one treats the Dirac mass term $m_{D}\nu_{L}\nu_{R}$ as a perturbative
term so that the Feynman diagram contains a mass insertion proportional
to $m_{D}$ and an intermediate $\nu_{L}$ propagator proportional
to $1/\slashed{p}$. The propagator effectively contributes a factor
of $1/m_{\nu_{R}}$ after imposing the on-shell condition of $\nu_{R}$.
So the diagram is suppressed by a factor of $\theta\approx m_{D}/m_{\nu_{R}}$,
and $C_{Z\to\nu_{L}+\nu_{R}}$ is suppressed by $\theta^{2}$. Therefore,
in either way, one obtains
\begin{align}
C_{Z\to\nu_{L}+\nu_{R}} & \approx\frac{y_{\nu}^{2}g^{2}m_{Z}^{3}T_{{\rm SM}}}{128\sqrt{2}\pi^{3}\cos^{2}\theta_{W}G_{F}m_{\nu_{R}}^{2}}K_{1}\left(\frac{m_{Z}}{T_{{\rm SM}}}\right),\\
C_{W\to e_{L}+\nu_{R}} & \approx\frac{y_{\nu}^{2}g^{2}m_{W}^{3}T_{{\rm SM}}}{64\sqrt{2}\pi^{3}G_{F}m_{\nu_{R}}^{2}}K_{1}\left(\frac{m_{W}}{T_{{\rm SM}}}\right),
\end{align}
where $g$ is the SM $SU(2)$ gauge coupling and $\theta_{W}$ is
the Weinberg angle. Compared to Eq.~\eqref{eq:-52}, the rates of
$\nu_{R}$ production via $Z$ and $W^{\pm}$ decays are enhanced
by $\sim1/(G_{F}m_{\nu_{R}}^{2})$. This conclusion is consistent
with the previous calculation in Ref.~\cite{Coy:2021sse}. 

Note that below the electroweak scale $\nu_{R}$ can also be produced
via neutrino oscillations which, as we will show below,  are much
more efficient than the decay processes. 

\subsection{Production rates of $\nu_{R}$ via oscillations\label{subsec:osc}}

Neutrino oscillations in the early universe have been extensively
studied as they play a key role  in sterile neutrino DM---see \cite{Dasgupta:2021ies,Kusenko:2009up,Abazajian:2017tcc,Adhikari:2016bei}
for recent reviews. In the non-resonant production (NRP) regime with
negligible lepton asymmetries, the production rate of $\nu_{R}$ reads~\cite{lesgourgues_book,Abazajian:2001nj,Gelmini:2019wfp}:
\begin{equation}
\Gamma_{\nu_{R}}\approx\theta_{{\rm eff}}^{2}\Gamma_{\nu_{L}}\thinspace,\label{eq:-50}
\end{equation}
with
\begin{align}
\Gamma_{\nu_{L}} & \approx1.27G_{F}^{2}\epsilon T_{{\rm SM}}^{5}\thinspace,\label{eq:-56}\\
\sin^{2}2\theta_{{\rm eff}} & \approx\frac{\sin^{2}2\theta}{\sin^{2}2\theta+(\cos2\theta-2\epsilon T_{{\rm SM}}V_{T}/m_{\nu_{R}}^{2})^{2}}\thinspace,\label{eq:-51}
\end{align}
where $\epsilon\approx3.15$ and $\theta_{{\rm eff}}$ is the effective
mixing angle which is related to but different from $\theta$ due
to the thermal MSW potential, $V_{T}\approx-10.88\times10^{-9}\ \text{GeV}^{-4}\times\epsilon T^{5}$.
Unlike the conventional MSW potential proportional to $G_{F}$, $V_{T}$
in the cosmological thermal plasma is proportional to $G_{F}^{2}$.
This is  because the contributions of particles and anti-particles
in the background cancel out at the leading order.   According to
Eq.~\eqref{eq:-51}, at low temperatures ($2\epsilon T_{{\rm SM}}|V_{T}|/m_{\nu_{R}}^{2}\ll\cos2\theta$),
we have $\theta_{{\rm eff}}\approx\theta$ while at high temperatures
$\sin^{2}2\theta_{{\rm eff}}$ would be suppressed by $T_{{\rm SM}}^{-12}$. 

By comparing $C_{{\rm osc.}}\equiv\Gamma_{\nu_{R}}n_{\nu_{L}}$ with
$\Gamma_{\nu_{R}}$ given in Eq.~\eqref{eq:-50} to Eq.~\eqref{eq:-52},
we obtain 
\begin{equation}
\frac{\int C_{h\to\nu_{L}+\nu_{R}}T_{{\rm SM}}^{-6}dT_{{\rm SM}}}{\int C_{{\rm osc.}}T_{{\rm SM}}^{-6}dT_{{\rm SM}}}\lesssim10^{-3}\frac{m_{\nu_{R}}}{1\ {\rm GeV}}\ll1\thinspace,\label{eq:-57}
\end{equation}
where as a crude approximation we have taken $\theta_{{\rm eff}}\approx\theta$
for $2\epsilon T_{{\rm SM}}|V_{T}|/m_{\nu_{R}}^{2}\leq10^{-1}$ and
only integrated $C_{{\rm osc.}}T_{{\rm SM}}^{-6}$ below the corresponding
temperature. The actual integral in the denominator is larger, leading
to a more suppressed ratio. Eq.~\eqref{eq:-57} implies that the production
of $\nu_{R}$ via neutrino oscillations is much more efficient than
that from Higgs decay, as long as $m_{\nu_{R}}$ is below the electroweak
scale. 

To check whether neutrino oscillations could bring $\nu_{R}$ into
thermal equilibrium, we compare $\Gamma_{\nu_{R}}$ with the Hubble
parameter:
\begin{equation}
\frac{\Gamma_{\nu_{R}}}{H}\sim10^{4}\gg1\thinspace,\label{eq:-58}
\end{equation}
where the ratio is estimated at the temperature when $\Gamma_{\nu_{R}}$
peaks, i.e., when $2\epsilon T_{{\rm SM}}V_{T}/m_{\nu_{R}}^{2}$ in
the denominator of Eq.~\eqref{eq:-51} is equal to $\cos2\theta\approx1$. 

Eq.~\eqref{eq:-58} implies that the production rate of $\nu_{R}$
via oscillation is sufficiently high to render $\nu_{R}$ thermal,
provided that its mass is below the electroweak scale and the mixing
angle $\theta\approx\sqrt{0.1\ \text{eV}/m_{\nu_{R}}}$ is determined
by the seesaw relation. This conclusion is consistent with Ref.~\cite{Dolgov:2003sg}
which obtained $m_{\nu_{R}}^{2}\sin^{4}2\theta\gtrsim10^{-5}\ \text{eV}^{2}$
as the condition for $\nu_{R}$ to reach equilibrium---see Eqs.~(38)
and (39) therein.

\subsection{Benchmarks}

As we have shown in Sec.~\ref{subsec:Yukawa} and Sec.~\ref{subsec:osc},
the production rate of $\nu_{R}$ in the early universe is sufficiently
high for $\nu_{R}$ to reach thermal equilibrium, provided that the
interaction of $\nu_{R}$ with the SM is determined by the type-I
seesaw mass relation\footnote{This is only true when we neglect the flavor structure, which in principle
could accommodate three rather hierarchical $\nu_{R}$'s. In fact,
one could have two $\nu_{R}$'s responsible for the observed light
neutrino masses and one $\nu_{R}$ with much more suppressed $y_{\nu}$
and $\theta$ than those given in Eqs.~\eqref{eq:yb} and \eqref{eq:-45}
so as to circumvent the restriction of the seesaw mass relation---see,
e.g., Refs.~\cite{Coy:2021sse,Coy:2022xfj}.}. Therefore, the type-I seesaw DM is either in the EE or EF case,
depending on the dark sector coupling $y$ introduced in Eq.~\eqref{eq:Yukawa}.

Now let us consider some specific benchmarks for the type-I seesaw
DM: 
\begin{align*}
\text{Benchmark L1: } & (m_{\nu_{R}},\ m_{\chi},\ m_{\phi})=(1,\ 10,\ 12)\ {\rm TeV\thinspace};\\
\text{Benchmark L2: } & (m_{\nu_{R}},\ m_{\chi},\ m_{\phi})=(1,\ 10,\ 12)\ {\rm GeV\thinspace};\\
\text{Benchmark H1: } & (m_{\nu_{R}},\ m_{\chi},\ m_{\phi})=(1,\ 0.1,\ 0.12)\ {\rm TeV\thinspace};\\
\text{Benchmark H2: } & (m_{\nu_{R}},\ m_{\chi},\ m_{\phi})=(1,\ 0.1,\ 0.12)\ {\rm GeV\thinspace}.
\end{align*}
Here L/H indicates that $\nu_{R}$ is much lighter/heavier than $\chi$.
Two of the benchmarks are at TeV scales and the other two at GeV scales.
We set a small but sizable mass splitting between $m_{\chi}$ and
$m_{\phi}$ so that eventually all $\phi$ particles produced in the
early universe decay to $\chi$ and other particles.

According to Tab.~\ref{tab:result1}, Benchmarks L1 and L2 would
be in the EE case if $y\gg10^{-4}$ or in the EF case if $y\ll10^{-4}$.
For Benchmarks H1 and H2, the criterion is similar except that $10^{-4}$
is changed to $10^{-7}$. 

\begin{figure}
\centering

\includegraphics[width=0.494\textwidth]{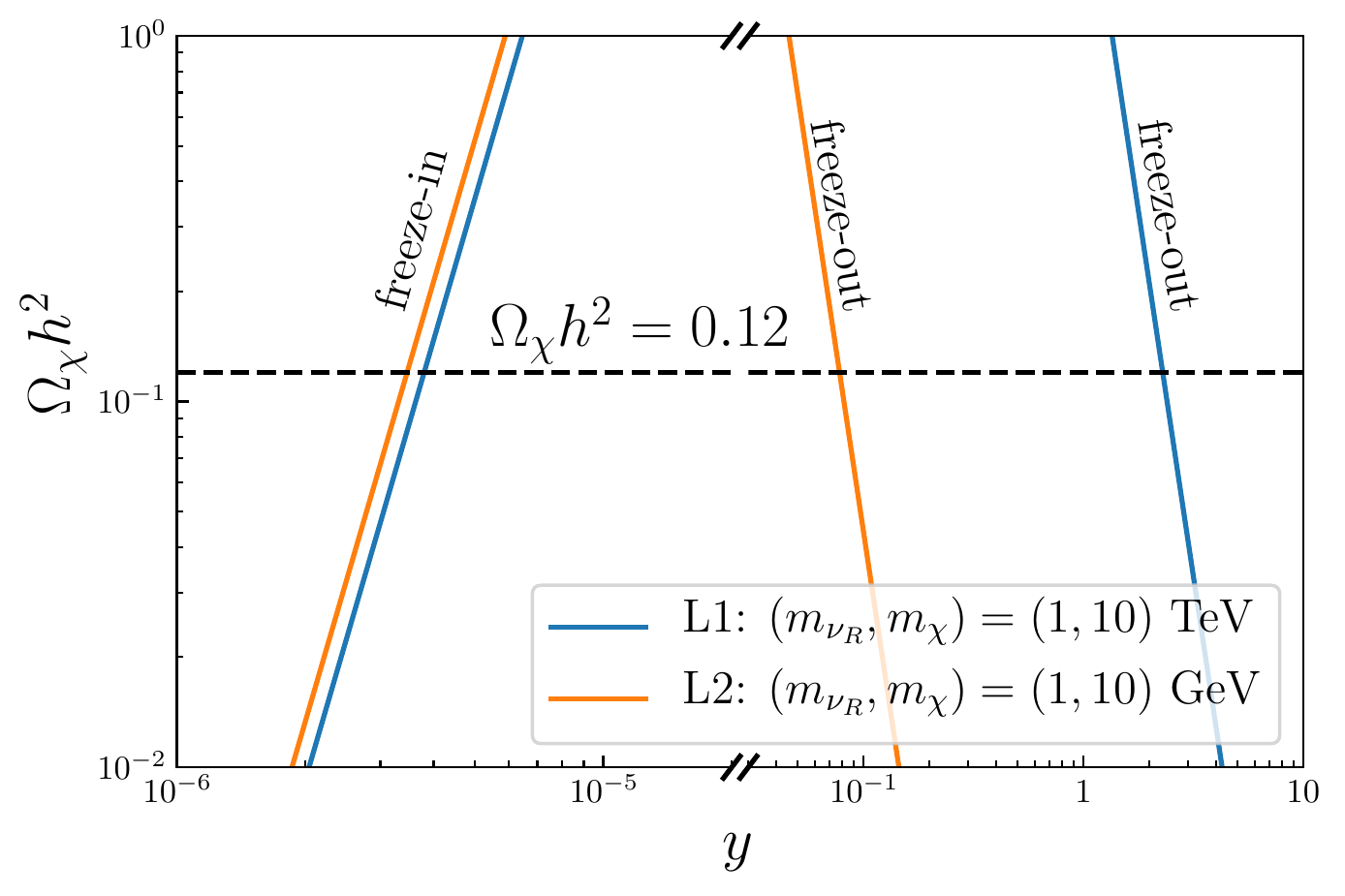}\includegraphics[width=0.49\textwidth]{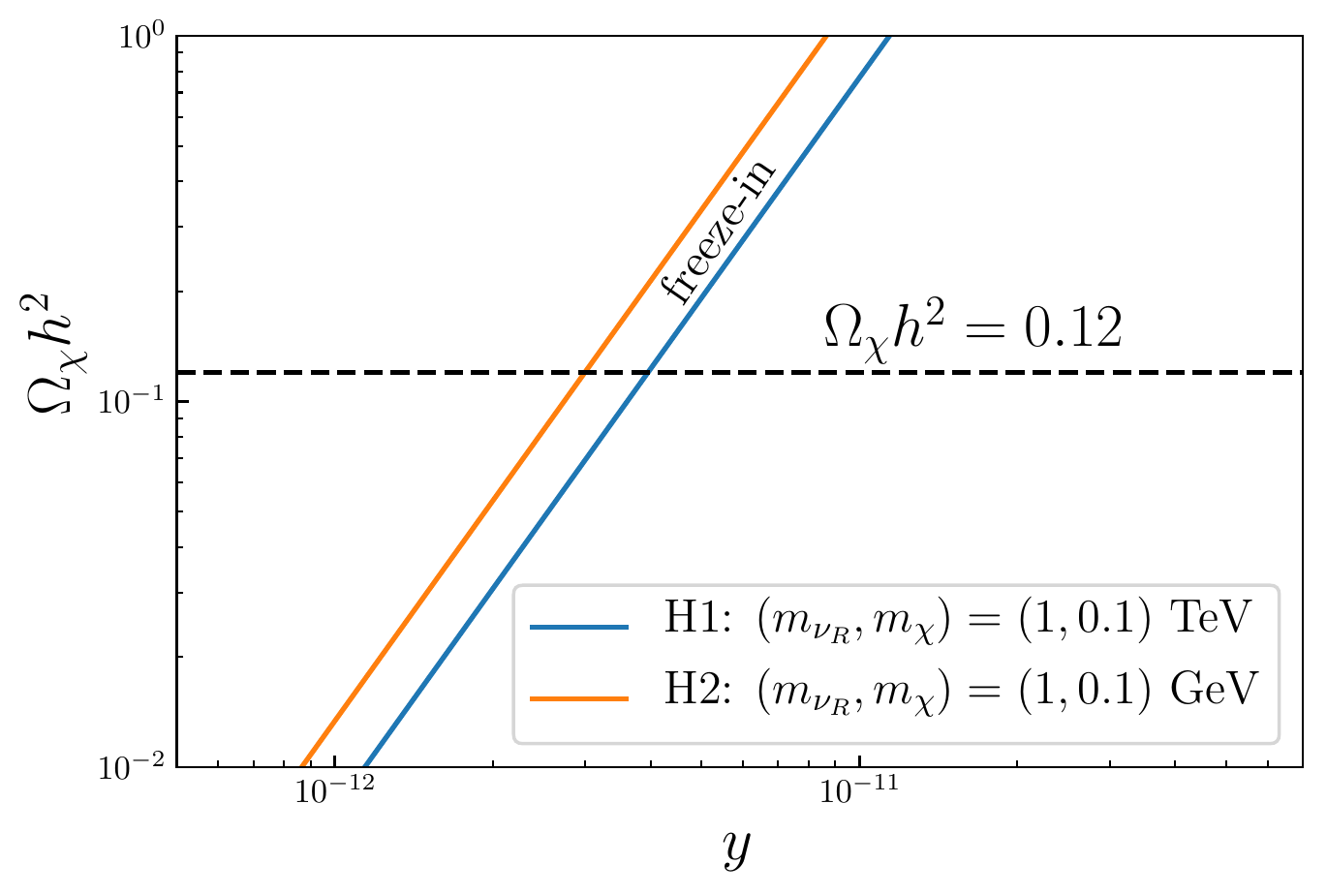}

\caption{The relic density of type-I seesaw DM for benchmarks L1/2 and H1/2.\label{fig:bench}
}
\end{figure}

Let us first concentrate on Benchmarks L1 and L2, for which the DM
candidate $\chi$ is produced via $2\nu_{R}\to2\chi$. When $y$ is
small ($y\ll10^{-4}$), this is a freeze-in process and, according
to Tab.~\ref{tab:result1}, we take Eq.~\eqref{eq:o-EFFF-1} with
$n_{\nu_{R}}/n_{\nu_{R}}^{{\rm eq}}=1$ to compute $\Omega_{\chi}h^{2}$.
 The results are presented in the left panel of Fig.~\ref{fig:bench}
as the increasing lines (labeled ``freeze-in'') which go across
$\Omega_{\chi}h^{2}=0.12$ at 
\begin{equation}
y=3.8\times10^{-6}\ (\text{for\ L1})\ \text{or}\ 3.5\times10^{-6}\ (\text{for\ L2})\thinspace.\label{eq:-59}
\end{equation}
The small difference is caused by the value of $g_{\star}$ at freeze-in
{[}see $g_{\star{\rm f.i.}}$ in Eq.~\eqref{eq:o-EFFF-1}{]} which
is evaluated at $T_{{\rm SM}}\sim m_{\chi}$. For L1, the freeze-in
temperature is well above the electroweak scale, hence $g_{\star{\rm f.i.}}\approx106.75$.
At $10$ GeV (the case of L2), $g_{\star{\rm f.i.}}$ falls to $83.51$~\cite{Wallisch:2018rzj}.

If we increase $y$ to a sufficiently large value, it will enter the
EE regime. The DM candidate $\chi$ keeps in thermal equilibrium via
$2\nu_{R}\leftrightarrow2\chi$ until it becomes non-relativistic
and freezes out from the thermal bath. According to Tab.~\ref{tab:result1},
the relic abundance $\Omega_{\chi}h^{2}$ is determined by Eq.~\eqref{eq:o-EEFE-1},
corresponding to the decreasing lines (labeled ``freeze-out'') in
the left panel of Fig.~\ref{fig:bench}. The two lines go across
$\Omega_{\chi}h^{2}=0.12$ at 
\begin{equation}
y=2.3\ (\text{for\ L1})\ \text{or}\ 0.078\ (\text{for\ L2})\thinspace.\label{eq:-60}
\end{equation}

Note that in the EE regime, larger $m_{\chi}$ implies larger $y$
in order to account for the correct DM relic abundance. If one sets
$m_{\chi}$ to higher values $\sim{\cal O}(100)$ TeV, then $y$ would
exceed the unitarity bound ($\sim\sqrt{4\pi}$). This is known as
the Griest-Kamionkowski bound, which dictates that thermal freeze-out
DM should be lighter than $340$ TeV~\cite{Griest:1989wd}, otherwise
the annihilation cross section would be too large and violate the
unitarity bound. 

Next, let us turn to Benchmarks H1 and H2. Due to $m_{\nu_{R}}\gg m_{\chi}$,
the DM candidate $\chi$ is directly produced via $\nu_{R}$ decay,
$\nu_{R}\to\chi\phi$. Therefore, the critical value of $y$ separating
the EE and EF cases is much smaller than that for L1 and L2. For $y\ll10^{-7}$,
Benchmarks H1 and H2 are in the EF regime. Hence we take Eq.~\eqref{eq:o-EF-2}
to compute $\Omega_{\chi}h^{2}$. The results are presented in the
right panel of Fig.~\ref{fig:bench} as the increasing lines which
go across $\Omega_{\chi}h^{2}=0.12$ at 
\begin{equation}
y=3.9\times10^{-12}\ (\text{for\ H1})\ \text{or}\ 3.0\times10^{-12}\ (\text{for\ H2})\thinspace.\label{eq:-59-1}
\end{equation}

For larger $y$, in principle it could enter the EE regime. However,
it turns that for Benchmarks H1 and H2 there are no freeze-out solutions
if the neutrino-Higgs coupling $y_{\nu}$ is determined by the seesaw
mass relation. This is because $\nu_{R}$ due to its heavy masses
in these two benchmarks cannot be the final states of DM (co-)annihilation
processes. The dark sector particles $\chi$ and $\phi$ have to annihilate
or co-annihilate to lighter SM species. Possible processes could be
$\chi\phi\to h\nu_{L}$ mediated by an off-shell $\nu_{R}$, $\chi\chi\to\nu_{L}\nu_{L}$
or $e_{L}e_{L}$ via box diagrams, $\chi\phi\to3\nu_{L}$ with $\nu_{R}$
and $Z$ as intermediate states, etc. If $\chi\phi\to h\nu_{L}$ is
not kinematically suppressed, it is the dominant process for DM freeze-out.
Using Eq.~\eqref{eq:o-EE-2} with $y'=y_{\nu}$ determined by Eq.~\eqref{eq:yb}
and neglecting the Higgs mass, one would obtain $y=2.3\times10^{5}$
for Benchmark H1. This obviously violates the unitarity bound. In
other words, if $y$ is limited within the unitarity bound ($y\lesssim\sqrt{4\pi}$),
then the cross section of the dominant freeze-out process would be
too small to deplete the overproduced DM.  For Benchmark H2 where
$m_{\chi}$ is well below the electroweak scale, $\chi\phi\to h\nu_{L}$
is kinematically suppressed. One would have to consider other processes
that are further suppressed by additional vertices in the diagrams. 

There is, however, an interesting scenario that could potentially
allow freeze-out solutions for $\nu_{R}$ heavier than $\chi$. According
to Eq.~\eqref{eq:o-EE-2} where $\Omega_{\chi}h^{2}$ is roughly {[}neglecting
all ${\cal O}(1\sim10)$ quantities{]} proportional to $m_{\nu_{R}}^{2}/y'^{2}=\frac{v^{2}}{2}m_{\nu_{R}}/m_{\nu_{L}}$,
if $\nu_{R}$ is sufficiently light (e.g., $\sim$keV), then $\Omega_{\chi}h^{2}$
could be suppressed by $m_{\nu_{R}}$, possibly leading to a freeze-out
solution. Using Eqs.~\eqref{eq:o-EE-2} and \eqref{eq:-14-2} with
$yy'=4\pi v^{-1}\sqrt{2m_{\nu_{L}}m_{\nu_{R}}}$ and $m_{\chi}/m_{\nu_{R}}=0.1$,
we find that the freeze-out solution is at $m_{\nu_{R}}\approx5$
keV. By varying $yy'$ and $m_{\chi}/m_{\nu_{R}}$ one can obtain
different values, but the mass scale cannot be much higher than the
keV scale. Despite the existence of such freeze-out solutions, this
scenario is ruled out by BBN observations due to additional thermal
species at the MeV scale. If we only consider that these particles
are well above the MeV scale, and if the magnitude of $y_{\nu}$ is
subject to the seesaw constraint, then the type-I seesaw DM with $\nu_{R}$
heavier than $\chi$ cannot have a freeze-out solution.

\begin{figure}
\centering

\includegraphics[width=0.49\textwidth]{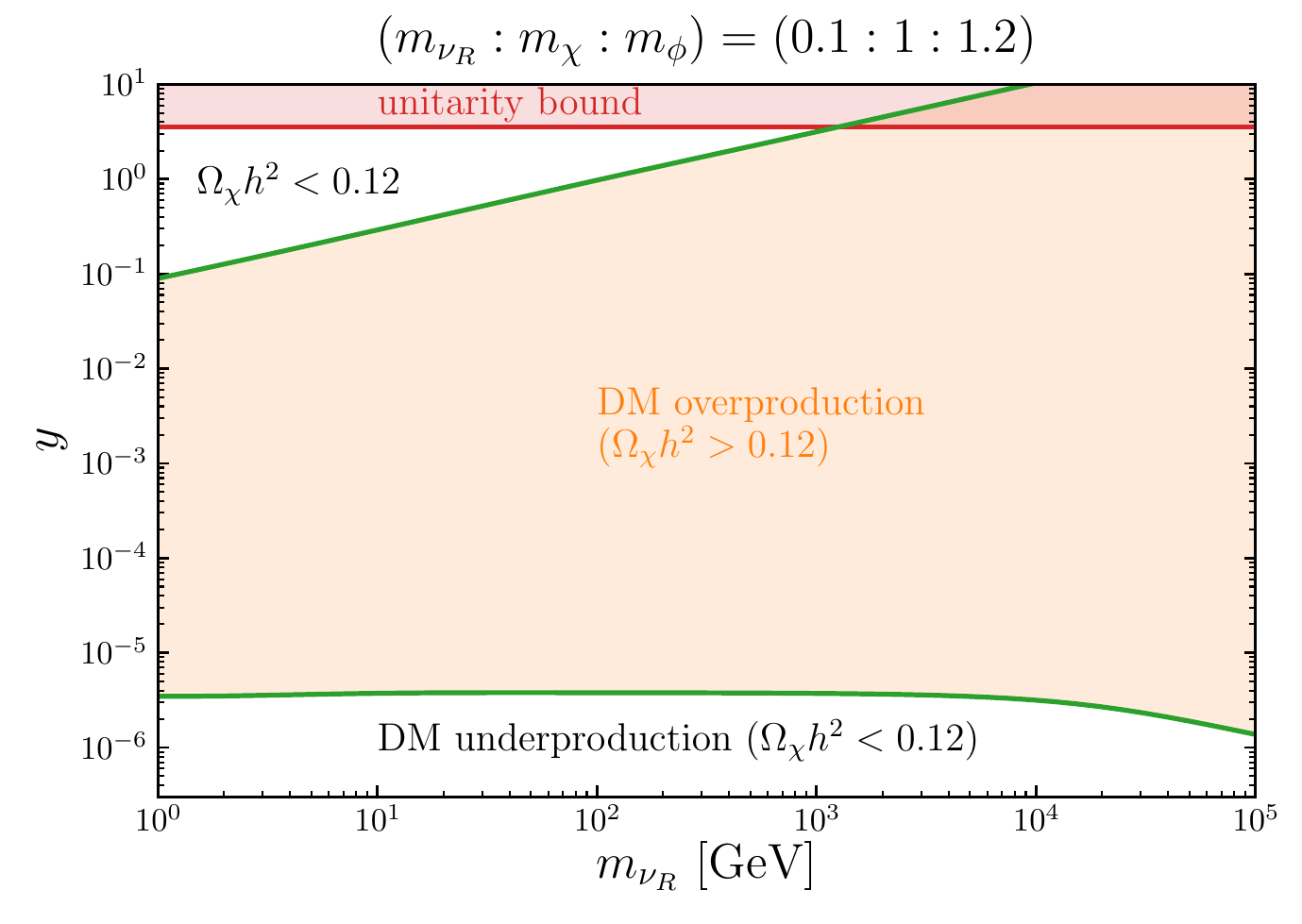}\includegraphics[width=0.494\textwidth]{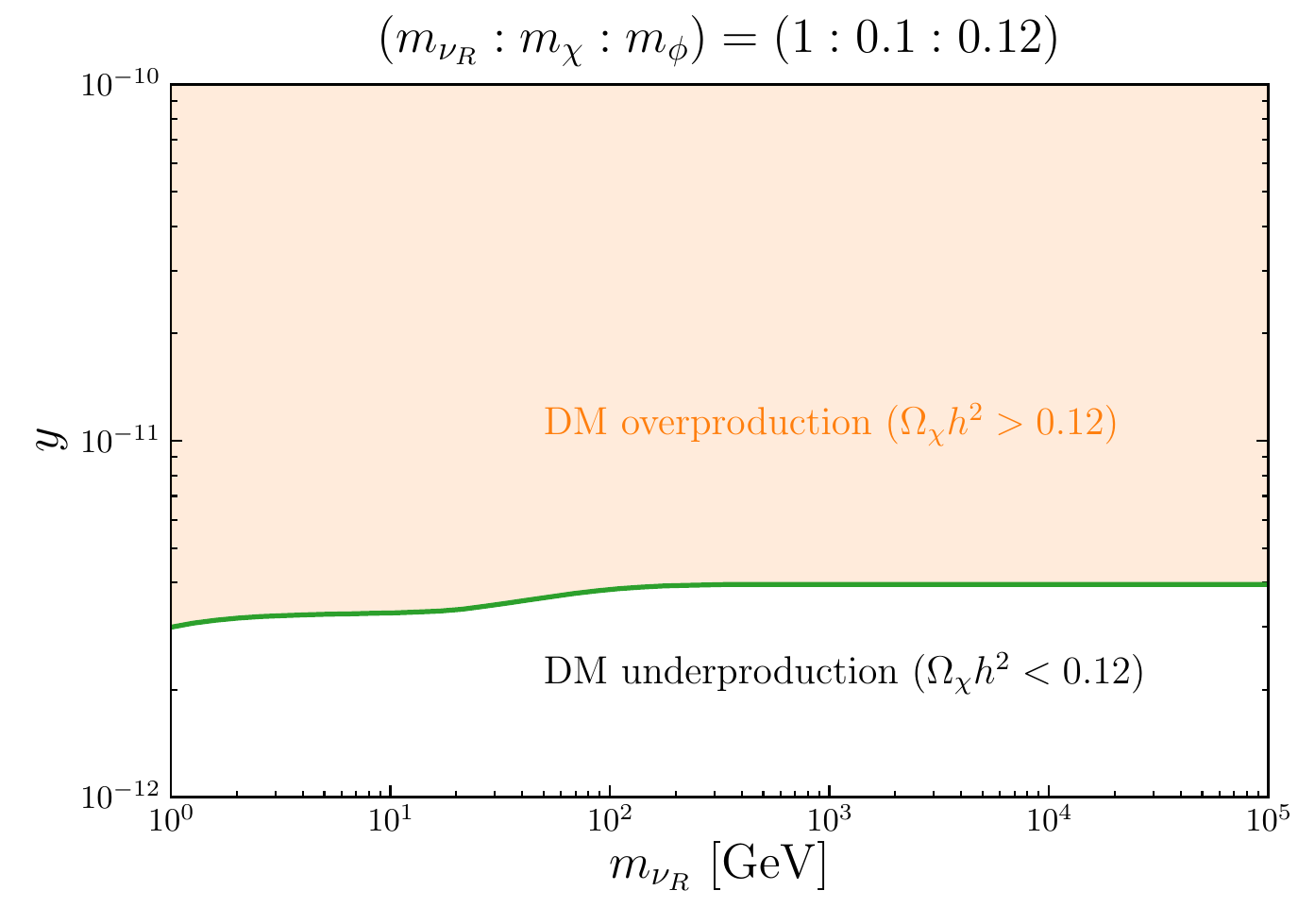}

\caption{The parameter space of the type-I seesaw DM model for two different
mass ratios, one with $m_{\nu_{R}}\ll m_{\chi,\phi}$ (left panel)
and the other with $m_{\nu_{R}}\gg m_{\chi,\phi}$ (right panel).
\label{fig:para-space} }
\end{figure}

In Fig.~\ref{fig:para-space}, we show the parameter space of the
type-I seesaw DM model assuming two different mass ratios, one with
$m_{\nu_{R}}\ll m_{\chi,\phi}$ and the other with $m_{\nu_{R}}\gg m_{\chi,\phi}$.
As previously discussed, the heavy $\nu_{R}$ scenario does not allow
for freeze-out solutions. So the blue curve in the right panel of
Fig.~\ref{fig:para-space} is a freeze-in solution for $\Omega_{\chi}h^{2}=0.12$.
The value of $y$ of this solution varies in a limited range from
$3\times10^{-12}$ to $4\times10^{-12}$ due to the variation of $g_{\star{\rm f.i.}}$.
In the left panel, the variation of the freeze-in solution due to
$g_{\star{\rm f.i.}}$ is insignificant but it bends down significantly
at $m_{\nu_{R}}\gtrsim10^{4}$ GeV because the $s$-channel contribution
in Eq.~\eqref{eq:o-EFFF-2} becomes dominant. In addition, there is
also a freeze-out solution (the top green line) which would exceed
the unitarity bound for $m_{\nu_{R}}\gtrsim1$ TeV.

\section{Observational consequences\label{sec:Observation}}

DM produced via the RHN portal is difficult to probe via direct detection
experiments because it does not directly interact with normal matter
consisting of electrons and quarks. Nevertheless, we would like to
point out an interesting process, $\chi+e_{L}^{-}\to\phi+\overline{\nu}_{L}+e_{L}^{-}$,
with $\nu_{R}$ and $W^{-}$ as intermediate states. This process
might be possible if the active-sterile neutrino mixing is not too
small. The major concern is that the non-relativistic $\chi$ might
not have sufficient energy to cause an observable electron recoil,
unless $\chi$ has been boosted.

Regarding indirect detection, the annihilation process $\chi\overline{\chi}\to\nu_{R}\overline{\nu_{R}}$
produces two RHNs which may subsequently decay to stable particles
($\gamma$, $e^{\pm}$, $\nu$, etc.) in the SM, as has recently been
studied comprehensively in Ref.~\cite{Morrison:2022zwk}.  Alternatively,
$e^{\pm}$ can be more straightforwardly produced from $\chi\overline{\chi}$
annihilation  at the one-loop level with $\phi$, $\nu_{R}$, and
$W^{\pm}$  appearing  in a box diagram. We make a crude estimation
and find that that the annihilation cross section is too small to
to be relevant to the current indirect detection experiments. In addition
to annihilation, one might consider the $\phi$ decay as a source
of indirect detection signals. As estimated below Eq.~\eqref{eq:-12},
if $\nu_{R}$ is light, $\phi\to\chi+\nu_{R}$ is possible but the
coupling or the mass splitting has to be extremely suppressed in order
to render $\phi$ long-lived at a cosmological time scale. If $\nu_{R}$
is heavy, $\phi$ may decay to $\chi$ and, via an off-shell $\nu_{R}$,
to some SM states. In this scenario, the lifetime of $\phi$ can be
substantially longer but whether there is viable parameter space for
observable indirect detection signals remains an open question. 

Perhaps the most interesting observational consequence is the cosmological
effective number of relativistic species, $N_{{\rm eff}}$. If RHNs
are sufficiently light, they may contribute to $N_{{\rm eff}}$. As
is well known, neutrinos can be Dirac particles and this possibility
motivates extensive studies on the potential contribution to $N_{{\rm eff}}$~\cite{Abazajian:2019oqj,Luo:2020sho,Borah:2020boy,Adshead:2020ekg,Luo:2020fdt}.
In our framework, a particularly noteworthy scenario is the FE case
with very light $\nu_{R}$. Through the freeze-in from the SM to $\nu_{R}$,
a considerably large amount of energy and entropy can be transferred
to the dark sector, which after undergoing freeze-out at a relatively
late epoch will release almost all the entropy stored in the dark
sector to $\nu_{R}$. In this scenario, the contribution to $N_{{\rm eff}}$
can be very significant without overproducing DM. In contrast, DM
frozen-in from $\nu_{R}$ or more directly from $\nu_{L}$ usually
cannot change $N_{{\rm eff}}$ significantly~\cite{Hufnagel:2021pso,Li:2021okx}
because the amount of energy transferred to DM is limited by $\Omega_{\chi}h^{2}=0.12$,
corresponding to $n_{\chi}=9.74\times10^{-12}\text{eV}^{-4}/(2m_{\chi})\approx6.3\times10^{-4}\text{cm}^{-3}\times(\text{MeV}/m_{\chi})$
which is much lower than the neutrino number density. The potentially
significant contribution to $N_{{\rm eff}}$ in the FE case calls
for a more dedicated study in light of current and upcoming  precision
measurements of $N_{{\rm eff}}$.

\section{Conclusion\label{sec:Conclusion}}

In this paper we have presented a comprehensive investigation of a
generic framework in which DM is produced from the RHN portal. As
formulated in Eq.~\eqref{eq:Yukawa}, our framework assumes that a
RHN ($\nu_{R}$) is coupled to a dark fermion ($\chi$, which serves
as DM with absolute stability) and a dark scalar ($\phi$). Since
the analyses crucially depend on  whether the equilibrium between
$\nu_{R}$ and the SM, and between the dark sector and $\nu_{R}$,
 can be established, we propose four basic cases, namely EE, EF, FF,
and FE, as illustrated in Fig.~\ref{fig:4-cases}.  Our results
of the DM relic density and the criteria for identifying the four
cases have been summarized in Tab.~\ref{tab:result1}. Note that
the results also depend on whether $\nu_{R}$ is heavy or light compared
the mass scale of the dark sector, because the dominant production
processes are decay and annihilation for heavy and light $\nu_{R}$,
respectively. 

As a simple application, we considered the type-I seesaw model extended
by the dark sector in our framework.  We found that when the neutrino-Higgs
Yukawa coupling is determined by the seesaw mass relation, the thermal
history of the dark sector can only be in the EE or EF case. For light
$\nu_{R}$, both have viable solutions accounting for the correct
DM relic abundance, while for heavy $\nu_{R}$, only EF is viable---see
Fig.~\ref{fig:bench}. 

Finally, we discussed a few observational consequences in this framework.
Although it seems difficult to cause observable signals in direct
detection experiments, the RHN-portal DM may have important implications
for indirect detection and future precision measurements of $N_{{\rm eff}}$. 

\begin{acknowledgments}
We  thank Rupert Coy for useful discussions. This work is supported
in part by the National Natural Science Foundation of China under
grant No. 12141501.
\end{acknowledgments}

\appendix

\section{Collision terms \label{sec:coll}}

The analytical calculations presented below assume the Maxwell-Boltzmann
statistics, which implies not only that the Maxwell-Boltzmann distribution
$f=e^{-E/T}$ is used for bosonic and fermionic species, but also
that $(1\pm f)$ in Eq.~\eqref{eq:-6} are neglected. With this assumption,
the number density can be computed as follows:
\begin{equation}
n=\int_{0}^{\infty}f(p)\frac{dp^{3}}{(2\pi)^{3}}=\int_{0}^{\infty}\exp\left(-\frac{\sqrt{p^{2}+m^{2}}}{T}\right)\frac{dp^{3}}{(2\pi)^{3}}=\frac{m^{2}T}{2\pi^{2}}K_{2}\left(\frac{m}{T}\right).\label{eq:-32}
\end{equation}
The collision terms for two-to-two, two-to-one, and one-to-two processes
 take the following forms:
\begin{align}
C_{1+2\rightarrow3+4} & =\int d\Pi_{1}d\Pi_{2}d\Pi_{3}d\Pi_{4}f_{1}f_{2}|{\cal M}|^{2}(2\pi)^{4}\delta^{4}\thinspace,\label{eq:-33}\\
C_{1+2\rightarrow3} & =\int d\Pi_{1}d\Pi_{2}d\Pi_{3}f_{1}f_{2}|{\cal M}|^{2}(2\pi)^{4}\delta^{4}\thinspace,\label{eq:-34}\\
C_{1\rightarrow3+4} & =\int d\Pi_{1}d\Pi_{3}d\Pi_{4}f_{1}|{\cal M}|^{2}(2\pi)^{4}\delta^{4}\thinspace.\label{eq:-35}
\end{align}
The collision terms computed assuming the Maxwell-Boltzmann statistics
typically deviate from the exact values by $\sim10\%$---see e.g.
Tab. III in Ref.~\cite{Luo:2020sho}.

\subsection{Two-to-two collision terms}

Let us first consider a two-to-two process, generically denoted by
$1+2\rightarrow3+4$. 

Assuming $m_{1}=m_{2}\equiv m$ and $f_{1,2}=e^{-E_{1,2}/T}$, the
collision term can be computed analytically~\cite{Gondolo:1990dk}:
\begin{equation}
C_{1+2\rightarrow3+4}\approx\frac{T}{32\pi^{4}}\int_{4m^{2}}^{\infty}s^{1/2}(s-4m^{2})\sigma_{1+2\to3+4}K_{1}\left(\frac{s^{1/2}}{T}\right)ds\thinspace,\label{eq:-30}
\end{equation}
where  $\sigma_{1+2\to3+4}$ is the total cross section of this process. 

For $\nu_{R}+\overline{\nu_{R}}\to\chi+\overline{\chi}$, the cross
section reads
\begin{equation}
\sigma_{2\nu_{R}\to2\chi}=\frac{y^{4}}{16\pi s}\sqrt{1-\frac{4m_{\chi}^{2}}{s}}+{\cal O}(\delta m^{2}/s)\thinspace,\label{eq:-31}
\end{equation}
where $\delta m^{2}\equiv m_{\phi}^{2}-m_{\chi}^{2}$. Substituting
Eq.~\eqref{eq:-31} into Eq.~\eqref{eq:-30}, we obtain the result
in Eq.~\eqref{eq:-10}. In particular, $C_{2\nu_{R}\to2\chi}$ has
the following high- and low-temperature limits (assuming $\nu_{R}$
in thermal equilibrium):
\begin{align}
\lim_{T\to\infty}C_{2\nu_{R}\to2\chi} & \approx\frac{y^{4}}{128\pi^{5}}T^{4}\thinspace,\label{eq:-36}\\
\lim_{T\to0}C_{2\nu_{R}\to2\chi} & \approx\frac{y^{4}}{256\pi^{4}}T^{3}m_{\chi}e^{-2m_{\chi}/T}\thinspace.\label{eq:-37}
\end{align}

For $\nu_{R}+\overline{\nu_{R}}\to\phi+\phi^{*}$, the cross section
reads
\begin{equation}
\sigma_{2\nu_{R}\to2\phi}=\frac{y^{4}}{8\pi s^{2}}\left[s\coth^{-1}\left(s/\overline{s}\right)-\overline{s}\right]+{\cal O}(\delta m^{2}/s)\thinspace,\label{eq:-38}
\end{equation}
where $\overline{s}\equiv\sqrt{s^{2}-4m_{\phi}^{2}s}$. Substituting
Eq.~\eqref{eq:-38} into Eq.~\eqref{eq:-30}, we find that the integral
of $s$ can only be calculated numerically but to a good approximation
($\sim10\%$ within $0.3<T/m_{\phi}<30$) it can be approximated as
\begin{equation}
C_{2\nu_{R}\to2\phi}\approx\alpha\log\left(\frac{T}{m_{\phi}}+e^{-\beta T/m_{\phi}}\right)C_{2\nu_{R}\to2\chi},\ \label{eq:-39}
\end{equation}
where $\alpha=1.77$ and $\beta=0.84$ are obtained via fitting. When
the collision term is used in Eq.~\eqref{eq:-9}, we are mainly concerned
with the integral $\int C_{2\nu_{R}\to2\phi}T^{-6}dT$. Numerically
performing the integral,  we find
\begin{equation}
\lim_{m_{\phi}\to m_{\chi}}\frac{\int_{0}^{\infty}C_{2\nu_{R}\to2\phi}T^{-6}dT}{\int_{0}^{\infty}C_{2\nu_{R}\to2\chi}T^{-6}dT}\approx1.87\thinspace,\label{eq:-40}
\end{equation}
which implies that with the same coupling $y$ and the small mass
splitting limit ($\delta m^{2}\ll m_{\chi}^{2}$),  the integrated
production rate of $\phi$ is almost twice as large as that of $\chi$.

For the co-annihilation process $\chi\phi\to BF$ appeared in Sec.~\ref{subsec:Case-EE-H},
we only consider cases with approximately equal masses of $\chi$
and $\phi$. In addition, the process is only important at $T\ll m_{\nu_{R}}$
so we assume that $m_{\nu_{R}}$ is sufficiently heavy in our calculation.
Under these assumptions, the cross section is given by
\begin{equation}
\sigma_{\chi\phi\to BF}=\frac{sy^{2}y'^{2}}{32\pi m_{\nu_{R}}^{2}\sqrt{s\left(s-4m_{\chi}^{2}\right)}}\thinspace.\label{eq:-61}
\end{equation}
Substituting Eq.~\eqref{eq:-61} into Eq.~\eqref{eq:-30}, we obtain
the corresponding collision term. For a general value of $T$, there
is no simple expression for the result. Technically one can express
the integral in terms of the Meijer G-function, but the expression
is not useful in practice. We are only interested in the approximate
result at $T\ll m_{\chi}$, for which we have 
\begin{equation}
C_{\chi\phi\to BF}\approx\frac{m_{\chi}^{3}T^{3}y^{2}y'^{2}}{128\pi^{4}m_{\nu_{R}}^{2}}e^{-2m_{\chi}/T}\thinspace.\label{eq:-62}
\end{equation}

\subsection{Two-to-one or one-to-two collision terms}

For $C_{1+2\rightarrow3}$ and $C_{1\rightarrow3+4}$ in Eqs.~\eqref{eq:-34}
and \eqref{eq:-35}, we assume the initial and final state masses
are negligible, respectively. Then we adopt the results from Appendix
A of Ref.~\cite{Luo:2020fdt} and obtain 
\begin{align}
C_{1+2\rightarrow3} & \approx\frac{|{\cal M}|^{2}}{32\pi^{3}}K_{1}\left(\frac{m_{3}}{T}\right)m_{3}T\thinspace,\label{eq:-64}\\
C_{1\rightarrow3+4} & \approx\frac{|{\cal M}|^{2}}{32\pi^{3}}K_{1}\left(\frac{m_{1}}{T}\right)m_{1}T\thinspace.\label{eq:-65}
\end{align}
It is worth mentioning that $C_{1\rightarrow3+4}$ has the following
high and low temperature limits:
\begin{align}
\lim_{T\to\infty}C_{1\rightarrow3+4} & \approx\frac{|{\cal M}|^{2}}{32\pi^{3}}T^{2}\thinspace,\label{eq:-63}\\
\lim_{T\to0}C_{1\rightarrow3+4} & \approx\frac{|{\cal M}|^{2}}{32\pi^{3}}m_{1}^{2}\sqrt{\frac{\pi}{2}}\left(\frac{T}{m_{1}}\right)^{3/2}e^{-m_{1}/T}\thinspace.\label{eq:-66}
\end{align}
For $C_{1+2\rightarrow3}$, the limits are the same, except that $m_{1}$
is replaced by $m_{3}$.

When dealing with the Boltzmann equation of a momentum distribution
function {[}see Eq.~\eqref{eq:f-boltz}{]}, one needs the following
collision terms:
\begin{align}
C_{1+2\rightarrow3}^{(f)} & =\frac{1}{2E_{3}}\int d\Pi_{1}d\Pi_{2}f_{1}f_{2}|{\cal M}|^{2}(2\pi)^{4}\delta^{4}\thinspace,\label{eq:-70}\\
C_{1\rightarrow3+4}^{(f)} & =\frac{1}{2E_{3}}\int d\Pi_{1}d\Pi_{4}f_{1}|{\cal M}|^{2}(2\pi)^{4}\delta^{4}\thinspace,\label{eq:-71}
\end{align}
where $3$ in $1+2\rightarrow3$ or $1\rightarrow3+4$ denotes the
particle of which the distribution is to be computed. 

Let us first compute Eq.~\eqref{eq:-71}, with the same assumptions
as aforementioned. Integrating out $d\Pi_{4}$, we obtain
\begin{equation}
C_{1\rightarrow3+4}^{(f)}=\frac{1}{2E_{3}}\int\frac{2\pi p_{1}^{2}dp_{1}dc_{13}}{(2\pi)^{3}2E_{1}}\frac{f_{1}|{\cal M}|^{2}}{2E_{4}}(2\pi)\delta(E_{1}-E_{3}-E_{4})|_{E_{4}\to\sqrt{p_{1}^{2}+p_{3}^{2}-2p_{1}p_{3}c_{13}}}\thinspace,\label{eq:-72}
\end{equation}
where $c_{13}\equiv(\mathbf{p}_{1}\cdot\mathbf{p}_{3})/(p_{1}p_{3})$.
One can further integrate out $c_{13}$ together with the $\delta$
function, provided that the argument in the $\delta$ function can
reach zero when $c_{13}$ scans from $-1$ to $1$. It is straightforward
to find out that this requires
\begin{equation}
p_{1}>p_{1}^{\min}\thinspace,\ \ p_{1}^{\min}=\left|\frac{m_{1}^{2}-4p_{3}^{2}}{4p_{3}}\right|\thinspace.\label{eq:-73}
\end{equation}
Integrating out $c_{13}$, we obtain
\begin{equation}
C_{1\rightarrow3+4}^{(f)}=\frac{1}{2E_{3}}\int_{p_{1}^{\min}}^{\infty}dp_{1}\frac{f_{1}|{\cal M}|^{2}p_{1}^{2}}{8\pi E_{1}(E_{1}-E_{3})}\frac{1}{\Delta}\thinspace,\label{eq:-74}
\end{equation}
where $\Delta=p_{1}p_{3}/|E_{1}-E_{3}|$ arsing from $\int\delta(E_{1}-E_{3}-E_{4})dc_{13}$.
Finally, integrating out $p_{1}$ with $f_{1}=e^{-E_{1}/T}$, we obtain
\begin{equation}
C_{1\rightarrow3+4}^{(f)}=\frac{T|{\cal M}|^{2}}{16\pi p_{3}^{2}}e^{-\frac{m_{1}^{2}+4p_{3}^{2}}{4p_{3}T}}\thinspace.\label{eq:-75}
\end{equation}
One can check that $\int C_{1\rightarrow3+4}^{(f)}d^{3}p_{3}/(2\pi)^{3}$
reproduces the result in Eq.~\eqref{eq:-65}.

As for $C_{1+2\rightarrow3}^{(f)}$, the calculation is similar except
that $p_{1}$ has both upper and lower bounds, $(E_{3}-p_{3})/2<p_{1}<(E_{3}+p_{3})/2$,
 and the result is given as follows:
\begin{equation}
C_{1+2\rightarrow3}^{(f)}=\frac{|{\cal M}|^{2}}{16\pi E_{3}}e^{-\frac{E_{3}}{T}}\thinspace.\label{eq:-76}
\end{equation}
Again, one can check that $\int C_{1+2\rightarrow3}^{(f)}d^{3}p_{3}/(2\pi)^{3}$
reproduces the result in Eq.~\eqref{eq:-64}.

\section{Calculation of the freeze-in momentum distribution\label{sec:app-f}}

The momentum distribution function, $f$, is governed by the following
Boltzmann equation:
\begin{equation}
\left[\frac{\partial}{\partial t}-Hp\frac{\partial}{\partial p}\right]f(t,\ p)=C^{(f)}\thinspace,\label{eq:f-boltz}
\end{equation}
where $C^{(f)}$ is the collision term for $f$. In the freeze-in
regime where the back-reaction is negligible, Eq.~\eqref{eq:f-boltz}
can be written as an integral of $C^{(f)}$. Below we show the details.

First, we introduce a transformation of variables from $(t,\ p)$
to $(a,\ x_{p})$ where $a$ is the scale factor ($\dot{a}/a=H$)
and $x_{p}=p/T$. Then the Jacobian matrix of this transformation
reads
\begin{equation}
\left(\begin{array}{c}
\frac{\partial}{\partial t}\\
\frac{\partial}{\partial p}
\end{array}\right)=\left(\begin{array}{cc}
Ha & Hx_{p}\\
0 & T^{-1}
\end{array}\right)\left(\begin{array}{c}
\frac{\partial}{\partial a}\\
\frac{\partial}{\partial x_{p}}
\end{array}\right).\label{eq:-77}
\end{equation}
Next, we express Eq.~\eqref{eq:f-boltz} in terms of $(a,\ x_{p})$.
As we will adopt $(a,\ x_{p})$ as new variables, we define
\begin{equation}
F(a,\ x_{p})=f(t,\ p)\thinspace,\label{eq:-78}
\end{equation}
and write
\begin{equation}
\left(1,\ -Hp\right)\left(\begin{array}{c}
\frac{\partial}{\partial t}\\
\frac{\partial}{\partial p}
\end{array}\right)f=\left(Ha,\ 0\right)\left(\begin{array}{c}
\frac{\partial}{\partial a}\\
\frac{\partial}{\partial x_{p}}
\end{array}\right)F\thinspace.\label{eq:-79}
\end{equation}
The zero entry in $\left(Ha,\ 0\right)$ above allows us to write
the left-hand side of Eq.~\eqref{eq:f-boltz} as a total derivative:
\begin{equation}
\frac{d}{da}F=\frac{C^{(f)}}{Ha}\thinspace.\label{eq:-80}
\end{equation}
Replacing $da\to dT=Ta^{-1}da$, we obtain
\begin{equation}
F(a,p/T)=\int_{T}^{\infty}\frac{C^{(f)}(T')}{H(T')T'}dT'\thinspace.\label{eq:-81}
\end{equation}

For a given collision term, one can use Eq.~\eqref{eq:-81} to compute
the momentum distribution function $f$. For example, taking the collision
term $C_{1\rightarrow3+4}^{(f)}$ in Eq.~\eqref{eq:-75}, we obtain
\begin{equation}
f=\frac{m_{{\rm pl}}e^{-x_{p}}|{\cal M}|^{2}}{8\pi g_{H}m_{1}^{3}x_{p}}\left(\sqrt{\pi x_{p}}\text{erf}\left(\frac{x_{m}}{2\sqrt{x_{p}}}\right)-ye^{-\frac{x_{m}^{2}}{4x_{p}}}\right),\label{eq:-82}
\end{equation}
where $x_{m}\equiv m_{1}/T$. In the limit of $x_{m}\gg\sqrt{x_{p}}$,
the ${\rm erf}$ function reduces to unity and Eq.~\eqref{eq:-82}
reduces to 
\begin{equation}
f=\frac{m_{{\rm pl}}|{\cal M}|^{2}}{8g_{H}m_{1}^{3}}\frac{1}{\sqrt{\pi x_{p}}}e^{-x_{p}}\thinspace.\label{eq:-83}
\end{equation}
From Eq.~\eqref{eq:-83}, one can readily obtain the average value
of $x_{p}$, $\langle x_{p}\rangle\equiv\int fx_{p}^{3}dx_{p}/\int fx_{p}^{2}dx_{p}=2.5$,
which is lower than the well-known values $\left(\langle x_{p}^{{\rm MB}}\rangle,\ \langle x_{p}^{{\rm FD}}\rangle,\ \langle x_{p}^{{\rm BE}}\rangle\right)=\left(3,\ 3.15,\ 2.70\right)$
for Maxwell-Boltzmann, Fermi-Dirac, and Bose-Einstein distributions,
respectively.

For two-to-two processes, the integral in Eq.~\eqref{eq:-81} generally
cannot be analytically integrated but the numerical evaluation is
straightforward. 

\section{Calculation of the secluded freeze-out\label{sec:freeze-out} }

In this appendix, we derive Eq.~\eqref{eq:-17} for the secluded freeze-out
scenario by revisiting the calculation for the standard freeze-out
mechanism, with some steps modified in order to take into account
that the dark sector has a different temperature as the SM sector.

Let us first inspect the freeze-out temperature, which is determined
by
\begin{equation}
n_{\chi}\langle\sigma v\rangle=H\thinspace,\label{eq:f}
\end{equation}
where $H$ approximately depends on $T_{{\rm SM}}$ while $n_{\chi}$
depends on the dark sector temperature $T_{\chi}$.  Before freeze-out,
$n_{\chi}$ in the non-relativistic regime is given by
\begin{equation}
n_{\chi}=\int e^{-E/T_{\chi}}\frac{d^{3}p}{(2\pi)^{3}}=\frac{m_{\chi}^{3}}{2\pi^{2}x}K_{2}\left(x\right)\approx e^{-x}m_{\chi}^{3}\left(\frac{1}{2\pi x}\right)^{3/2}\thinspace.\label{eq:f-1}
\end{equation}
where $x$ has been defined in Eq.~\eqref{eq:x}. 

The freeze-out value of $x$, denoted by $x_{{\rm f.o.}}$, is determined
by substituting Eq.~\eqref{eq:f-1} and Eq.~\eqref{eq:-7} into Eq.~\eqref{eq:f}:
\begin{equation}
e^{-x_{{\rm f.o}}}=\frac{\left(2\pi\right)^{3/2}g_{H}}{\epsilon_{{\rm f.o}}^{2}m_{\chi}m_{{\rm pl}}x_{{\rm f.o}}^{1/2}\langle\sigma v\rangle}\thinspace,\label{eq:f-3}
\end{equation}
which can be solved numerically to determined $x_{{\rm f.o}}$.

For the DM annihilation process, $2\chi\to2\nu_{R}$, $\langle\sigma v\rangle$
is given by Eq.~\eqref{eq:-15}:
\begin{equation}
\langle\sigma v\rangle\approx\frac{y^{4}}{32\pi}\left[\frac{K_{1}\left(x_{{\rm f.o}}\right)}{m_{\chi}K_{2}\left(x_{{\rm f.o}}\right)}\right]^{2}\approx\frac{y^{4}}{32\pi m_{\chi}^{2}}\left[1-3x_{{\rm f.o}}^{-1}+6x_{{\rm f.o}}^{-2}+{\cal O}(x_{{\rm f.o}}^{-3})\right].\label{eq:-16-2}
\end{equation}
Using Eq.~\eqref{eq:-16-2}, we find that the solution of $x_{{\rm f.o.}}$
in Eq.~\eqref{eq:f-3} can be approximately fitted  by
\begin{equation}
x_{{\rm f.o}}=35.5+9.52\log_{10}\left[y\cdot\epsilon_{{\rm f.o.}}^{1/2}\cdot\left(\frac{m_{\chi}}{{\rm GeV}}\frac{g_{H}}{17.2}\right)^{-1/4}\right].\label{eq:-14-a}
\end{equation}

For the DM co-annihilation process, $\chi\phi\to BF$, we have $\langle\sigma v\rangle=C_{\chi\phi\to BF}/(n_{\chi}n_{\phi})$
with $C_{\chi\phi\to BF}$ given by Eq.~\eqref{eq:-62}. The approximate
solution reads:
\begin{equation}
x_{{\rm f.o}}\approx22.2+4.88\log_{10}\left[yy'\cdot\left(\frac{m_{\chi}}{\text{GeV}}\right)^{1/2}\cdot\left(\frac{m_{\nu_{R}}}{1\ {\rm TeV}}\right)^{-1}\cdot\left(\frac{g_{H}}{17.2}\right)^{-1/2}\right].\label{eq:-14-2-a}
\end{equation}

\begin{figure}[h]
\centering

\includegraphics[width=0.6\textwidth]{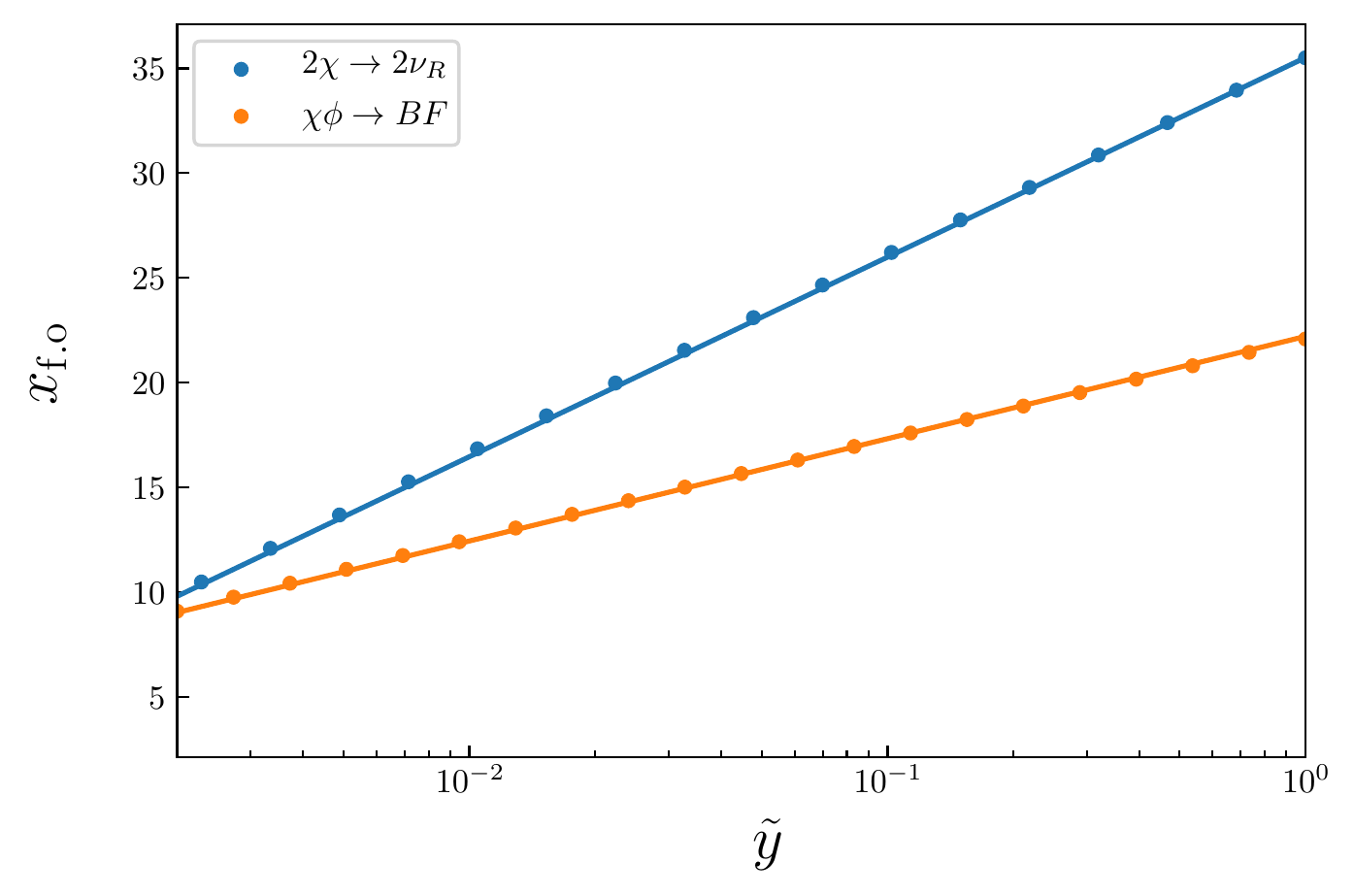}\caption{\label{fig:x-f-o} The approximate solutions (solid lines) for $x_{{\rm f.o}}$
compared to the exact ones (dots). The blue and orange solid lines
are obtained using Eqs.~\eqref{eq:-14-a} and ~\eqref{eq:-14-2-a}
respectively. }
\end{figure}

In Fig.~\ref{fig:x-f-o}, we show how well the above approximate
expressions can fit the exact results. In the figure, $\tilde{y}$
denotes the quantities in the square brackets of Eqs.~\eqref{eq:-14-a}
and ~\eqref{eq:-14-2-a}.

Once $x_{{\rm f.o}}$ is obtained from solving Eq.~\eqref{eq:f-3},
the number density at the moment of freeze-out can be computed by
\begin{equation}
n_{\chi{\rm f.o.}}=\frac{m_{\chi}^{2}g_{H{\rm f.o.}}}{m_{{\rm pl}}\langle\sigma v\rangle x_{{\rm f.o}}^{2}\epsilon_{{\rm f.o.}}^{2}}\thinspace.\label{eq:-24}
\end{equation}

After freeze-out, the comoving number density $n_{\chi}a^{3}$ is
conserved while in the SM sector, many species will eventually annihilate
and inject their energy into lighter species. Nevertheless, the SM
comoving entropy density $s_{{\rm SM}}a^{3}$ is conserved.  Hence
the ratio $n_{\chi}/s_{{\rm SM}}$ after freeze-out is a constant:
\begin{equation}
\frac{n_{\chi}}{s_{{\rm SM}}}=\left.\frac{n_{\chi}}{s_{{\rm SM}}}\right|_{{\rm f.o.}}.\label{eq:f-2}
\end{equation}
Using Eqs.~\eqref{eq:f-2}, \eqref{eq:-24}, \eqref{eq:-7},  and
$s_{{\rm SM}}=2\pi^{2}g_{\star}^{(s)}T_{{\rm SM}}^{3}/45$, we obtain
\begin{equation}
n_{\chi}=\frac{2\pi^{3/2}\epsilon x_{{\rm f.o.}}g_{\star}^{(s)}T_{\text{SM}}^{3}}{3\sqrt{5}m_{\chi}m_{{\rm pl}}\langle\sigma v\rangle\sqrt{g_{{\rm \star f.o}}}}\thinspace.\label{eq:-25}
\end{equation}
 Eq.~\eqref{eq:-25} remains valid even in relatively late eras dominated
by matter or by vacuum energy, provided that $s_{{\rm SM}}$ in these
eras are interpreted as the entropy density of radiation (photons
and neutrinos) and $T_{\text{SM}}$ in Eq.~\eqref{eq:-25} is interpreted
as the photon temperature. Although the universe is dominated by non-radiation
content in these eras, the comoving entropy of photons and neutrinos
is conserved. Taking $g_{\star}^{(s)}=3.938$ and $T_{\text{SM}}=2.7$
K, we obtain from Eq.~\eqref{eq:-25} today's energy density of DM:
\begin{equation}
\left.\rho_{\chi+\overline{\chi}}\right|_{{\rm today}}=2m_{\chi}n_{\chi}=1.35\times10^{-11}\ \text{eV}^{4}\cdot\frac{x_{{\rm f.o.}}\epsilon}{\sqrt{g_{{\rm \star f.o}}}}\cdot\frac{10^{-9}\ \text{GeV}^{-2}}{\langle\sigma v\rangle}\thinspace.\label{eq:-26}
\end{equation}
It is conventional to write  the result in terms of $\Omega_{{\rm DM}}h^{2}$
which is defined by
\begin{equation}
\Omega_{{\rm DM}}\equiv\frac{\rho_{\text{DM}}}{\rho_{{\rm cri.}}}\thinspace,\ \rho_{{\rm cri.}}\equiv\frac{3H_{0}^{2}m_{{\rm pl}}^{2}}{8\pi}\thinspace,\ h\equiv\frac{H_{0}}{100\ \text{km}/\text{sec}/\text{Mpc}}\thinspace,\label{eq:-27}
\end{equation}
where  $H_{0}$ is the Hubble constant today and $\rho_{{\rm cri.}}$
is the critical energy density. Note that $\Omega_{{\rm DM}}h^{2}$
is independent of $h$ (or $H_{0}$) so that it is not affected by
the long-standing Hubble tension problem---see \cite{DiValentino:2021izs,Efstathiou:2021ocp,Schoneberg:2021qvd}
for latest reviews. Combining Eqs.~\eqref{eq:-26} and \eqref{eq:-27},
we obtain
\begin{equation}
\frac{\Omega_{{\rm DM}}h^{2}}{0.12}=\frac{\rho_{\text{DM}}}{9.74\times10^{-12}\ {\rm eV}^{4}}=\frac{x_{{\rm f.o.}}\epsilon}{\sqrt{g_{{\rm \star f.o}}}}\cdot\frac{1.4\times10^{-9}\ \text{GeV}^{-2}}{\langle\sigma v\rangle}\thinspace.\label{eq:-28}
\end{equation}

\bibliographystyle{JHEP}
\bibliography{AllRef}

\end{document}